\documentclass{llncs}

\usepackage{latexsym}
\usepackage{amsfonts}
\usepackage{amsmath}
\usepackage{amssymb}
\usepackage{color}
\usepackage{xspace}
\usepackage{graphicx}
\usepackage{subfigure}
\usepackage{cite}
\usepackage{balance}
\usepackage{apacite}
\usepackage{wrapfig}
\usepackage{multirow}
\usepackage{url}
\usepackage{epstopdf}
\usepackage{epsfig}
\usepackage{geometry}

\usepackage{algorithm}
\usepackage{algorithmicx}
\usepackage[noend]{algpseudocode}

\newcommand{\eat}[1]{}

\newcommand{\ie}{i.e.,\xspace}
\newcommand{\eg}{e.g.,\xspace}

\newcommand{\stitle}[1]{\vspace{1.5ex}\noindent{\bf #1}}
\newcommand{\sstab}{\rule{0pt}{8pt}\\[-1.8ex]}

\newcommand{\etitle}[1]{\vspace{0.8ex}\noindent{\underline{\em #1}}}
\newcommand{\img}{\textsc{Img}\xspace}

\newcommand{\bi}{\begin{itemize}}
\newcommand{\ei}{\end{itemize}}

\newcommand{\supp}{Sup}

\newcommand{\mni}{\textsc{MnIS}\xspace}
\newcommand{\frq}{\kw{frq}}

\newcommand{\myhrule}{\rule[.5pt]{\hsize}{.5pt}}
\newcommand{\mat}[2]{{\begin{tabbing}\hspace{#1}\=\+\kill #2\end{tabbing}}}

        {\end{itemize}} 

\newcounter{prop}
\renewcommand{\theprop}{\arabic{theorem}}
\newenvironment{prop}{\begin{em}
        \refstepcounter{theorem}
        {\vspace{1.5ex}\noindent \bf Proposition \theprop:}}{
        \end{em}\eop\vspace{1.5ex}}

\newcommand{\kw}[1]{{\ensuremath {\mathsf{#1}}}\xspace}

\newcounter{ccc}
\newcommand{\bcc}{\setcounter{ccc}{1}\theccc.}
\newcommand{\icc}{\addtocounter{ccc}{1}\theccc.}

\newcommand{\eop}{\hspace*{\fill}\mbox{$\Box$}}     

\newcommand{\aka}{\emph{a.k.a.}\xspace}

\newcommand{\ptnmine}{\kw{DisMiner}}

\newcommand{\topksch}{\textsc{ETSearch}\xspace}
\newcommand{\topkpm}{\textsc{TopKPM}\xspace}
\newcommand{\fpm}{{FPM}\xspace}
\newcommand{\naive}{\textsc{Naive}\xspace}
\newcommand{\grami}{{GRAMI}\xspace}
\newcommand{\agrami}{{AGRAMI}\xspace}

\newcommand{\apm}{\textsc{AprTopK}\xspace}
\newcommand{\wrt}{\emph{w.r.t.}\xspace}

\newcommand{\nopt}{\kw{Nopt}}
\newcommand{\upruning}{\kw{UPruning}}
\newcommand{\rpick}{\kw{Rpick}}
\newcommand{\sverify}{\kw{Sverify}}
\newcommand{\nrst}{\kw{Nrst}}
\newcommand{\mk}{\kw{Mkst}}
\newcommand{\uplevel}{\textsc{UplevelPruning}}
\newcommand{\forwardtree}{\textsc{FwTreeGen}\xspace}
\newcommand{\backwardtree}{\textsc{BwTreeGen}\xspace}
\newcommand{\lu}{\kw{L_{u}}}
\newcommand{\te}{\textsc{FrqChk}\xspace}
\newcommand{\traverse}{\textsc{Traverse}\xspace}
\newcommand{\aprtopk}{\textsc{AprTopK}\xspace}
\usepackage{enumerate}

\begin{document}



\title{Near-optimal Top-$k$ Pattern Mining}

\author{
Xin~Wang\inst{a} (xinwang@swpu.edu.cn),
Zhuo~Lan\inst{a} (202022000326@stu.swpu.edu.cn), 
Yu-Ang~He\inst{a} (202022000311@stu.swpu.edu.cn),
Yang~Wang\inst{a} (wangyang@swpu.edu.cn),
Zhi-Gui~Liu\inst{b} (liuzhigui@swust.edu.cn),
Wen-Bo~Xie\inst{a,}\thanks{Corresponding author at: School of Computer Science, Southwest Petroleum University, Chengdu 610500, China. E-mail: wenboxie@swpu.edu.cn (Wen-Bo Xie)} (wenboxie@swpu.edu.cn)
}

\institute{
$^a$ School of Computer Science, Southwest Petroleum University,  \\Chengdu 610500, China \\
$^b$ School of Information Engineering, Southwest University of Science and Technology, \\ Mianyang 621010, China \looseness=-1
}


\maketitle

\vspace{-2ex}
\begin{abstract}
Nowadays, frequent pattern mining (FPM) on large graphs receives increasing attention, since it is crucial to a variety of applications, e.g., social analysis. Informally, the FPM problem is defined as finding all the patterns in a large graph with frequency above a user-defined threshold. However, this problem is nontrivial due to the unaffordable computational and space costs in the mining process. In light of this, we propose a cost-effective approach to mining near-optimal top-$k$ patterns. Our approach applies a ``level-wise'' strategy to incrementally detect frequent patterns, hence is able to terminate as soon as top-$k$ patterns are discovered. Moreover, we develop a technique to compute the lower bound of support with smart traverse strategy and compact data structures. Extensive experimental studies on real-life and synthetic graphs show that our approach performs well, i.e., it outperforms traditional counterparts in efficiency, memory footprint, recall and scalability. \looseness=-1

\eat{
(\fpm) problem aims at identifying all the subgraphs (\aka patterns) that appear frequently in a graph according to a user-defined frequency threshold. The problem is of critical to many applications, such as social network analysis. While, on large scale graphs, the \fpm problem becomes even 
The \fpm problem on large scale graph, however, is a tough issue for classical methods which require great computational and memory costs. Therefore, an matches-storage-free method, called \te, is proposed in this paper. In the \te method, the Isomorphism Check problem is transformed into the Info Constraint Satisfaction problem. Thus the calculation of the support in this method does not reply on matches stored in memory, instead, only the information of domains and structures are considered, resulting in a great decrease in memory cost. Then, based on this method, an approximate algorithm is developed to efficiently detect top-$k$ patterns, which reserves the early termination property and avoids the costly enumeration of all matches. Extensive tests on real-life and synthetic graphs experimentally verify that our method is of validity. Meanwhile, the efficiency and recall of our method outperform other traditional counterparts.}

\end{abstract}

\section{Introduction}
\label{sec-intro}

Frequent pattern mining is one of the most important problems in knowledge discovery and graph mining, of which the main task is to find subgraphs with support above a threshold, from a dataset. There are two main types of settings considered to detect frequent patterns in previous researches, i.e., transactional-based and single-graph-based.
Recently, the single-graph-based setting has given rise to a high degree of academic attention, owing to its wide applications in e.g., bioinformatics \cite{bioinformatics2019}, cheminformatics \cite{MedicinalChemistry2021}, web analysis and social network analysis \cite{SocialNetworks2020}. 
Methods that rely on the single-graph-based setting mostly follow the combinatorial pattern enumeration paradigm. 
However, it is costly and unnecessary to enumerate all the patterns in real-world applications such as social network analysis \cite{huan2004spin, YanH03}. \looseness=-1

The minimum-image-based support (\mni for short) \cite{MNI} is widely used in traditional FPM algorithms due to its simplicity of calculation.
Generally, the traditional algorithms maintain all the matches of a pattern to calculate its \mni support. This brings big challenges to the mining evaluation on large graphs, as there may exist (potentially) exponentially many matches of a pattern in a large graph, which leads to an unsatiable memory cost and low scalability. \looseness=-1


\eat{
In the big data era, the scale of the frequent pattern is on the increase as well as the number of items in the pattern. The methods based on all the stored matches are limited, while the matches-storage-free methods that have good scalability will almost certainly take over them.
}


In addition to the scalability, the practicability is also considerable. In most real-world applications, it is unnecessary to enumerate all the patterns. On one hand, people prefer to focus on some typical patterns rather than scan the dazzling low-value ones \cite{zhu2011mining}. On the other hand, given a frequent pattern, all of its sub-patterns must be frequent as well, thus these sub-patterns are to some extent considered as ``redundant'' patterns. 

\eat{These highlight the need for {\em approximate top-$k$ pattern mining}: given a single large graph $G$, a support threshold $\theta$ and an integer $k$ that is used to find a given number of patterns, that not only satisfy support constraint but also are most interesting for the users. Furthermore, if an algorithm for the problem preserves the {\em early termination property}, i.e., it discovers top-$k$ patterns without identifying the entire pattern set, then we do not have to pay the price of costly pattern mining.}

\vspace{-2ex}
\begin{figure}
\centering
\includegraphics[width=\linewidth, keepaspectratio]{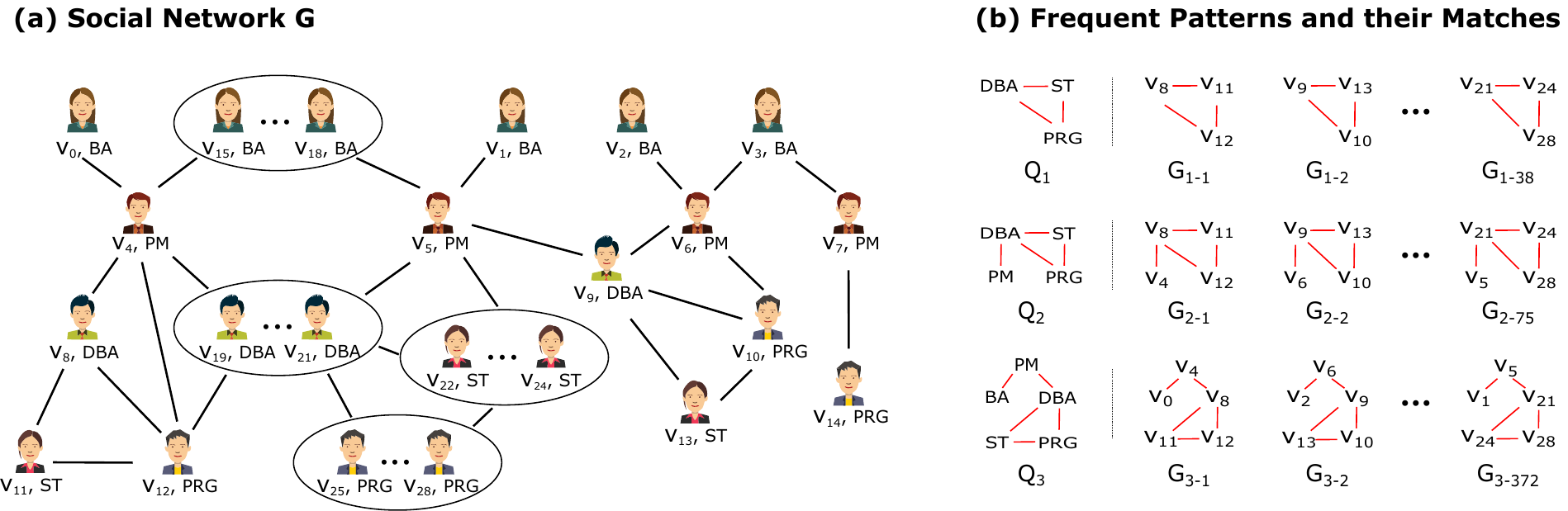}
\caption{ A snapshot of a social graph $G$ \& three patterns along with their matches}
\label{fig:example1}
\vspace{-4ex}
\end{figure}

\eat{
\begin{figure}
\centering
\includegraphics[width=3in, keepaspectratio]{./fig/example_intro2-1.eps}
\caption{a sequence of patterns}
\label{fig:example2}
\end{figure}

\begin{example}
\label{exa-graphs2}
Fig.~\ref{fig:example1} shows a sequence of patterns, among which the latter pattern is extended from the previous pattern. Obviously patterns $Q_0$, $Q_1$, $Q_2$, and $Q_3$ are subgraphs of pattern $Q_4$. If we return pattern $Q_4$, then there is no need to return to pattern $Q_0$, $Q_1, Q_2, Q_3$, since if a graph is frequent, then all its subgraphs are frequent.
\eop
\end{example}
}

\eat{The two examples call for techniques for {\em approximate top-$k$ pattern mining} with an efficient representation of matches. }

\eat{
\begin{example}
A fraction of a social graph $G_1$ is shown in Fig.~\ref{fig:example1} (a), where each
node denotes a person with an identifier consisting of a tag (e.g., celebrity (\kw{C}), learning blogger (\kw{LB}), entertainment blogger (\kw{EB}), student (\kw{S})) along with a subscript; and each edge indicates who follows whom, e.g., (\kw{S_0}, \kw{C_0}) indicates that \kw{S_0} follows \kw{C_0}. To simply layout, some nodes with the same tag are shrunk as a whole. From graph $G$, one may find that (i) there exist quite a few subgraphs as matches of pattern graphs $Q_1$ and $Q_2$ (Fig.~\ref{fig:example1} (b)); (ii) all the subgraphs (e.g., $Q_1$) of $Q_2$ have matches no less than $Q_2$, if they are all considered frequent and returned, we will have to face a particularly large set of frequent patterns, which not only includes ``redundancy'' but also is costly for inspection. Instead, we only need top-$k$ patterns. Then the cost for inspection and mining can be greatly reduced. For example, when $k=1$, $Q_2$ is considered more interesting than $Q_1$ from the perspective of {\em closeness}~\cite{YanH03} and hence is more preferred.  \eop 
\end{example}
}

\begin{example}
\label{exa-graphs}
A fraction of a social graph $G$ is shown in Fig.~\ref{fig:example1} (a), where each node denotes a person with ID and job title (e.g., project manager (PM), database administrator (DBA), programmer (PRG), business analyst (BA) and software tester (ST)); and each edge indicates friendship, e.g., ($v_0$, $v_4$) indicates that $v_0$ and $v_4$ are friends. 
From graph $G$, one can discover a few typical patterns, e.g., $Q_1$, $Q_2$ and $Q_3$ as well as their matches (Fig.~\ref{fig:example1} (b)). Note that $Q_1$ and $Q_2$ are both the subgraphs of $Q_3$, if they are considered frequent and returned, then we will have to face a large set of frequent patterns, which not only includes ``redundancy'' but also is costly for inspection. Instead, we only need top-$k$ patterns. Then the cost for inspection and mining can be greatly reduced. For example, when $k=1$, $Q_3$ is considered more interesting than $Q_1$ and $Q_2$ from the perspective of  closeness~\cite{YanH03} and hence is more preferred. \looseness=-1
\eop
\eat{
find that (i) there exist three frequent patterns  $Q_1$, $Q_2$ and $Q_3$, along with their matches (Fig.~\ref{fig:example1} (b)); (ii) $Q_1$ and $Q_2$ are both the subgraphs of $Q_3$, if they are considered frequent and returned, then we will have to face a large set of frequent patterns, which not only includes ``redundancy'' but also is costly for inspection. Instead, we only need top-$k$ patterns. Then the cost for inspection and mining can be greatly reduced. For example, when $k=1$, $Q_3$ is considered more interesting than $Q_1$ and $Q_2$ from the perspective of  closeness~\cite{YanH03} and hence is more preferred. \looseness=-1
\eop
}
\end{example}

The example suggests us to investigate {\em top-$k$ pattern mining} problem. While two crucial questions have to be answered:

(1) What metrics for measuring support and interestingness of a pattern shall we choose?

(2) How to develop an efficient algorithm such that (i) mining computation can terminate as soon as $k$ patterns are identified and (ii) support evaluation can be processed less costly in both evaluation time and memory footprint?

\eat{
The above requirements call for techniques for {\em approximate top-$k$ pattern mining} with the efficient representation of matches.
To tackle this issues, several questions have to be answered. (1) How to calculate \mni without storing matches?  (2) What metric can be easily used to measure the goodness of a pattern?
(3) How to develop an effective method such that mining computation can terminate as soon as $k$ ``best'' patterns have been identified? \looseness=-1
}

\stitle{Contributions.} This paper investigates the {\em  top-$k$ pattern mining} problem, and provides an effective approach to mining {\em near-optimal} top-$k$ patterns. Our contributions are as follows. \looseness=-1

(1) We adopt minimum-image-based support and propose a metric for measuring ``interestingness'' of a pattern. Based on the metrics, we formalize the {\em top-$k$ pattern mining} (\textsc{TopkPM}) problem and show the intractability of the problem (Section~\ref{sec-pre}).

\eat{
novel method, called \te, to calculate \mni without storing all the matches memory (Section~\ref{sec-te}). Instead, this method dynamically calculates the domain of sub-patterns by traversing each association space in the \kw{Domains} of the corresponding parent pattern in a depth-first manner.
}

(2) We investigate the \textsc{TopkPM} problem and develop an approach to identifying {\em near-optimal} top-$k$ patterns. The algorithm has following desirable performances: (a) it preserves {\em early termination property}, hence can terminate as soon as $k$ preferred patterns are discovered; and (b) the pattern set shows high recall value, compared with the optimal solution via intensive tests (Section~\ref{sec-alg}). 

(3) To facilitate support evaluation, we devise a novel technique for fast estimation.
Our technique, which captures the essential feature of \mni-based metric, well plugs into our main algorithm that works in a ``level-wise'' manner, hence is able to estimate the \mni support efficiently and accurately, while consuming  much less memory space (Section~\ref{sec-te}). 

(4) Using real-life and synthetic graphs, we experimentally verify the performances of our algorithm and find the following  (Section~\ref{sec-expt}). (a) Our algorithm shows excellent performance \wrt response time and memory cost on various real-life graphs. In particular, the required response time of our algorithm is about one order of magnitude faster than its counterparts. 
(b) Our algorithm, though incorporates approximation scheme, is able to obtain desired recalls, i.e., the set of top-$k$ patterns identified is {\em near-optimal}. For example, on two real-life graphs, our algorithm even achieves 100\% recall. (c) Our algorithm scales much better than its counterparts, \wrt response time and memory footprint. 

\eat{All the proofs and complementary experimental studies can be found in~\cite{full}. }


\section{Related Work}

The FPM problem on single large graphs has been well studied and a host of techniques have been proposed. We next review them as follows. \looseness=-1

\eat{
\etitle{Transaction-based}. Algorithms for pattern mining in graph
databases are given in~\cite{inokuchi2000apriori,huan2004spin}. The algorithms can be categorized into two types: (i) apriori based methods~\cite{inokuchi2000apriori}, and (ii) pattern-growth methods~\cite{huan2004spin}.
}

\stitle{Exact mining.} A large part of prior works focus on mining exact results. On static graphs, \cite{GraMi2014} formulated the FPM as a constrained satisfaction problem, and proposed an efficient algorithm called GraMI. \cite{0001QGW20} divided the workload by prefix projection to achieve efficient frequent pattern mining on multicore machines.
A framework \cite{ARAFF2021} was proposed to effectively reduce the duplicate and enormous frequent patterns through the initiation of a new ranking measurement called FSP-Rank.
On weighted graphs, \cite{AshrafHIALMW19, LeVNFL20} proposed approaches to detecting frequent patterns with weights. Over evolving graphs, \cite{AbdelhamidCSBCK17} introduced another dynamic algorithm IncGM+, which divides an input graph into frequent and infrequent updated subgraphs and prunes the update area by adjusting the boundary subgraphs named ``fringe''.  This approach keeps small memory overhead.
To tackle the distributive FPM problem and leverage parallel computation, DISTGRAPH \cite{TalukderZ16} uses a set of optimizations and efficient collective communication operations to minimize the total amount of messages shipped among different sites.
ScaleMine \cite{147877139} leverages the approximate and exact phases to achieve better load balance and more efficient evaluation when mining candidate patterns.
\cite{ESMFD2018} adopts a message-passing-free scheme among workers and utilizes a task scheduler to dynamically balance the workload for frequent subgraph mining on distributed systems. 
For the methods with depth-first order, gSpan \cite{yan2002gspan} designs a DFS lexicographic order to support the mining algorithm. FFSM \cite{2003Efficient} develops a new graph canonical form and completely avoids subgraph isomorphism testing by maintaining an embedding set for each frequent subgraph. Gaston \cite{2004A} adopts a step-wise approach that uses combinations of frequent paths, frequent free trees, and cyclic graphs to discover frequent subgraphs. \looseness=-1

\vspace{-0.3ex}
\stitle{Approximate mining.} To support practical applications, a host of techniques were developed for approximate pattern mining, under various settings. In \cite{GraMi2014}, an approximate solution called AGRAMI was also proposed to produce an incomplete set of frequent patterns with no false positives. 
On graphs with noise, exact matching is no longer feasible, \cite{82020Mining} introduced an approach, which allows inexact matching, to mining frequent patterns. 
Sampling-based algorithms have been proposed for the issue. 
\cite{TIPTAP2021} presented TipTap, a collection of sampling-based approximation algorithms for mining frequent $k$-vertex patterns in fully-dynamic graphs.
\cite{MaNIACS2021} proposed another sampling-based randomized algorithm called MaNIACS, of which the accuracy can be guaranteed by empirical Vapnik-Chervonenkis (VC) dimension.
\cite{LGSM2021} introduced a graph sampling algorithm RASI to reduce the unessential structure of a data graph. RASI demonstrates higher efficiency and greater accuracy than its counterparts for \textsc{FPM}.
REAFUM \cite{92015REAFUM} focuses on finding non-redundant representative frequent patterns that summarize the frequent patterns using approximate matching in a graph database. 
APGM \cite{15An} models the noise distribution through a probability matrix, and then uses an approximate matching strategy to mine useful patterns from the noise map database.
VEAM \cite{16ACOSTAMENDOZA2012381} mines frequent subgraphs under the semantic of inexact matching. The approach identifies frequent patterns from a collection of images with slight angular differences between the positions of image segments. On uncertain graphs, \cite{172019Efficient} developed an approximation algorithm with accuracy guarantee for the FPM problem under probabilistic semantic. 

\eat{RAM \cite{182008RAM} based on feature retrieval, constructs a vector of hash values for features and mines frequent graph patterns in a depth first fashion.}

\stitle{Top-$k$ mining.} The topic of identifying $k$ best patterns arose much attention in recent years. 
\cite{TopkDurableGPQueries2019} proposed an algorithm for mining top-$k$ durable matches in dynamic graphs, which uses a compact representation of the graph snapshots and appropriate time indexes to prune the search space. 
\cite{ourDTopK2021} proposed a metric to measure the quality of a pattern and developed a parallel algorithm with early termination property to efficiently discover $k$ best patterns in a distributed large graph. 
FastPat framework \cite{FastCore-based2021} utilizes the meta index and an upper bound of the frequency score to prune unqualified candidates. In particular, FastPat efficiently calculates the support of candidates through a join-based approach. 
\cite{CoTopK2020} uses a holistic best-first exploration strategy along with a compressed data structure called Replica to identify pairs of subgraph patterns that frequently co-occur in proximity within a single graph.
\textsc{Resling} \cite{RESLING2018} is a framework to mine the top-$k$ representative patterns. It evaluates patterns from the edit map and performs diversified ranking through two random-walk-based algorithms.
\cite{AslayNMG18} addressed the problem of approximate $k$-vertex frequent pattern mining on a dynamic graph with high probability in a given time.
To mine the top-$k$ uncertain frequent patterns from uncertain databases, \cite{LeVHNB20} introduced an approach that combines the mining and ranking phases as a whole to improve efficiency and reduce the memory cost.

\stitle{Our work differs from earlier works in two main aspects:} (1) a ``level-wise'' strategy is employed in the mining process to ensure the {\em early termination property}; (2) a novel support evaluation technique, that leverages wise traversal strategy and compact data structures is incorporated in the mining process. As a result, our method is committed to delivering  {\em near-optimal} results (the recall is up to 100\%) with low computational and memory costs. \looseness=-1

\eat{
Our results can guarantee that each pattern is a frequent pattern of the single large graph, but usually incomplete (with recall about 80\%~90\%).
}

\newcommand{\smlr}{\prec_{E,T}}


\section{Graphs, Patterns and Top-$k$ Pattern Mining}
\label{sec-pre}

In this section, we first review graphs, patterns, graph pattern matching; we then formalize the {\em top-$k$ pattern mining} problem. 


\subsection{Graph Pattern Matching}
\label{sec-patterns}

\begin{definition} {\bf Graph \& Subgraph.} A data graph (or simple graph) is defined as $G = (V, E, L)$,
where (1) $V$ is a set of nodes; (2) $E\subseteq V \times V$ is a set of undirected edges; and (3) each node $v\in V$ 
carries a tuple $L(v)=(A_1=a_1,A_2=a_2,\cdots,A_n=a_n)$, in which $A_i=a_i (i\in [1,n])$ represents that the node $v$ has a value $a_i$ for the attribute $A_i$, and is denoted as $v.A_i=a_i$. 

A graph $G_s = (V_s, E_s, L_s)$ is a
{\em subgraph} of $G = (V, E, L)$,
denoted by $G_s \subseteq G$,  if $V_s \subseteq V$, $E_s \subseteq E$, and
moreover, 
for each $v \in V_s$, $L_s(v) = L(v)$.
\eop
\end{definition} 


\eat{
A wide range of graph models e.g., property graphs, 
can be captured as heterogeneous data graphs.
where each node can be a person, a program or an attribute value of
another node,
and edges represent various relationships, e.g., create, knows, likes.
}

\begin{definition} {\bf Pattern \& Sub-pattern.} A { pattern} $Q$ is defined as a graph $(V_p, E_p, f_v)$, where $V_p$ and $E_p$ are the set of nodes and 
edges, respectively;
for each $u$ in $V_p$, 
it is associated with a predicate $f_v(u)$ defined as a conjunction of atomic formulas of the form of $`A=a$' such that $A$ denotes an attribute of the node $u$ and $a$ is a value of $A$. Intuitively, $f_v(u)$ specifies search conditions imposed by $u$, that is, for a node $v$ in $G$, if for each atomic formula $`A=a$' in $f_v(u)$, there is an attribute $A$ in $L(v)$ with $v.A=a$, then the node $v$ satisfies $f_v(u)$ (denoted as $v\sim u$). 

A pattern $Q' = (V_p', E_p', f_v')$ is subsumed by another pattern $Q = (V_p, E_p, f_v)$, denoted by $Q' \sqsubseteq Q$, if $(V_p', E_p')$ is a subgraph of $(V_p, E_p)$, and function $f_v'$ is a restriction of $f_v$. Then, $Q'$ is referred to as a sub-pattern of $Q$ if $Q' \sqsubseteq Q$. 
\eop
\end{definition}


\begin{definition}
\label{defGPM}
{\bf Pattern Matching}. We adopt the subgraph isomorphism~{\rm \cite{isomorphism}} as the matching semantic. A subgraph $G_s$ of $G$ matches a pattern $Q$ via isomorphism, iff there exists a bijective function $\rho$: $V_s\Rightarrow V_p$, such that (i) for each $v\in V_s$, $v\sim \rho(v)$ and (ii) $(v_i,v_j)\in E_s$ iff $(\rho(v_i),\rho(v_j))\in E_p$. 
\end{definition}

In a graph $G$, if there exists a subgraph $G_s$ that is mapped from $Q$ via $\rho$, 
then $G_s$ is referred to as a {\em match} of $Q$ in $G$, and the match set $M(Q,G)$ includes all the matches $G_s$ of $Q$ in $G$. Abusing the notation of {\em match}, we denote $v$ in $G_s$ as a {\em match} of $u$ in $Q$, if $\rho(u)=v$. Then for each node $u$ in $E_p$, one can derive a set $\{v | v\in G_s, G_s\in M(Q, G), v\sim u\}$ from $M(Q, G)$, and denote it by $\img(u)$. One may verify that $\img(u)$ consists of a set of {\em distinct} nodes $v$ in $G$ as {\em matches} of $u$ in $Q$. \looseness=-1

\eat{
When an isomorphism $\rho$ from a pattern $Q$ to a subgraph $G_s$ of $G$ exists, we say $G$ matches $Q$, and denote $G_s$ as a match of $Q$ in $G$. Abusing notations, we say $v$ in $G_s$ as a {\em match} of $u$ in $Q$, when $\rho(u)=v$. We use $M(Q, G)$ to denote the set of matches $G_s$ (resp. $\rho(Q)$) of $Q$ in $G$. 
Then, for each node $u$ in $E_p$, we derive a set $\{v | v\in \rho(Q), \rho(Q)\in M(Q, G), v=\rho(u)\}$ from match set $M(Q, G)$, and denote it by $\img(u)$. Intuitively, 
$\img(u)$ contains a set of {\em distinct} nodes $v$ in $G$ as matches of $u$ in $Q$.
}

\eat{
\begin{definition} 
\label{defGPM}
{\bf Graph Pattern Matching}. 
Consider graph $G$ and pattern $Q$, a node $v$ in $G$ satisfies the search conditions of a node $u$ in $Q$, denoted as $v\sim u$, if for each atomic formula $`A=a$' in $f_v(u)$, there exists an attribute $A$ in $L(v)$ such that $v.A=a$. Following the semantic of subgraph isomorphism~{\rm \cite{isomorphism}}, a match of pattern $Q$ in graph $G$ is a bijective function $\rho$ from the nodes of $Q$ to the nodes of a subgraph $G$, such that \looseness=-1

(1) for each node $u \in V_p$, $\rho(u)\sim u$, and;

(2) $(u, u')$ is an edge in $Q$ if and only if
$(\rho(u), \rho(u'))$ is an edge in $G$.\\ 
When an isomorphism $\rho$ from pattern $Q$ to a subgraph $G_s$ of $G$ exists, we say $G$ matches $Q$, and denote $G_s$ as a match of $Q$ in $G$. Abusing notations, we say $v$ in $G_s$ as a {\em match} of $u$ in $Q$, when $\rho(u)=v$. 
\eop
\end{definition}
}

\eat{
To ease presentation, we use $M(Q, G)$ to denote the set of matches $G_s$ (resp. $\rho(Q)$) of $Q$ in $G$. 
Then, for each node $u$ in $E_p$, we derive a set $\{v | v\in \rho(Q), \rho(Q)\in M(Q, G), v=\rho(u)\}$ from match set $M(Q, G)$, and denote it by $\img(u)$. Intuitively, 
$\img(u)$ contains a set of {\em distinct} nodes $v$ in $G$ as matches of $u$ in $Q$.
}

\eat{
\begin{example}
\label{exa-1}
Recall Fig.~\ref{fig:example1} (a).
Given patterns $Q_1$, $Q_2$ and $Q_s$ in Fig.~\ref{fig:example1} (b), one may verify that $Q_1$, $Q_2$ are {\em subsumed by $Q_3$}, and moreover, $M(Q_1, G)=\{G_{11}, G_{12}, G_{13}, G_{14}\}$, $M(Q_2, G)=\{G_{21}, G_{22}, G_{23}, G_{24}\}$ and  $M(Q_3, G)=\{G_{31}, G_{32}, G_{33}\}$. 
\end{example}
}

\begin{definition} 
\label{def-expansion} {\bf Forward \& Backward Expansions}. Given a pattern $Q$, its DFS tree $T_Q$ can be built via
a depth-first search on $Q$ from one of its node $u$. 
Then, edges in $T_Q$ are referred to as forward edges and the remaining edges in $Q$ are denoted as backward edges. Thus, the Forward expansion enlarges $Q$ by including a new edge from an existing node in $Q$ to a newly introduced node; while the Backward Expansion includes a new edge from two existing nodes of $Q$. \looseness=-1
\eop
\end{definition} 

\vspace{-1ex}
For example, a pattern $Q_c$ with edge set $\{({\rm ST},{\rm DBA}),({\rm DBA},{\rm PRG})\}$ can be generated via {\em forward expansion} from a pattern with edge $({\rm ST},{\rm DBA})$; with $Q_c$, another pattern $Q_1$ (shown in Fig.~\ref{fig:example1}(b)) is generated via {\em backward expansion}.

\eat{
Note that {\em similarity predicates} can be used instead of
equality ``='', e.g., string similarity, with no impact on our
algorithms. We consider ``='' to simplify the discussion.
}
\eat{
Note that the equality condition ``='' can be specified by
 similarity and equality models, such as string similarity,
or transformation functions~\cite{yang2014schemaless}.
}


\eat{
For each pattern node $u$, we use
$Q(u, G)$ to denote the set of all {\em matches} of $u$ in $Q(G)$,
i.e., $Q(u, G)$ consists of nodes $v$ in $G$ such that there exists a
function $h$ under which a subgraph $G' \in Q(G)$ is isomorphic
to $Q$, $v \in G'$ and $h(u) = v$.
}

\eat{
\begin{figure}[t]
\vspace{-1ex}
\centerline{\includegraphics[scale=0.36]{./fig/r-example.eps}}
\vspace{-2ex}
\centering
\caption{Labeled social graphs}
\label{fig-graphs}
\vspace{-3.5ex}
\end{figure}
}

\eat{
\begin{example}
\label{exa-dis}
As shown in Fig.~\ref{fig-rules} (a), a fragmentation of $G$
is $(F_1, F_2, F_3)$, where $F_1$, $F_2$ and $F_3$ are stored in sites
$S_1$, $S_2$ and $S_3$, respectively. Take $F_2$ as example, $F_2.O$ consists
of virtual nodes \texttt{Mat}, \texttt{Mary} and \texttt{Tim}, and the crossing edges are $(\texttt{Walt}, \texttt{Mat})$, $(\texttt{Fred}, \texttt{Mary})$, and $(\texttt{Fred}, \texttt{Tim})$.
To discovery the subgraph with node set $\{\texttt{Walt}, \texttt{Nancy}, \texttt{Mat}, \texttt{Dan}, \texttt{Emmy}\}$ that can match $Q_l$, data has to be shipped between $S_1$ and $S_2$. \eop
\end{example}
}

\eat{Observe that if $Q' \sqsubseteq Q$, then for any graph $G'$
that matches $Q$,
there exists a subgraph $G''$ of $G'$
such that $G''$ matches $Q'$. 
}

\stitle{Other Notations.} (1) The total size $|G|$ of $G$ (resp. $|Q|$ of $Q$) is $|V|+|E|$ (resp. $|V_p|+|E_p|$), i.e., the total number of nodes and edges in $G$ (resp. $Q$).
(2) For a pattern $Q$, its {\em complete} pattern $\widehat{Q}$ is such a pattern that takes the same set of nodes as $Q$, and moreover, has an edge for each pair of nodes in $\widehat{Q}$. 
(3) The height of a node $v$ in a rooted and directed tree ${\cal T}$ is the length of the longest path from $v$ to a leaf node of ${\cal T}$. Similarly, the height $h$ of ${\cal T}$ is the maximum height among all nodes in ${\cal T}$. 

A summary of notations are listed in Table~\ref{tab-terms}. 



\subsection{Top-$k$ Pattern Mining Problem}
\label{sec-fpm}

Below, we first review the frequent pattern mining problem, and then formalize the {\em top-$k$ pattern mining} (\textsc{TopkPM}) problem. We start from the support metric. \looseness=-1 

\begin{definition}
{\bf Support.} The support of a pattern $Q$ in
a single graph $G$, denoted by $\supp(Q, G)$, indicates the appearance
frequency of $Q$ in $G$. 
\eop
\end{definition}

Analogous to the association rules for itemsets,
the support metric for patterns should be anti-monotonic, i.e., for patterns $Q$ and $Q'$, if $Q'\sqsubseteq Q$, then
$\supp(Q', G) \geq \supp(Q, G)$ for any $G$, to facilitate search space pruning. Various pattern-based anti-monotonic support metrics exist, e.g.,
{\bf M}i{\bf n}imum-{\bf I}mage-based {\bf S}upport (\mni) \cite{MNI}, harmful overlap \cite{FiedlerB07} and maximum independent sets \cite{GudesSV06}. In this paper, \mni is chosen as the support metric owing to the merit of fast evaluation. 

Formally, the metric is defined as, 
\begin{equation}
    Sup(Q, G) = \min \left\{ |\img(u)| \mid u\in V_p \right\}, 
\end{equation}
\eat{
\begin{equation}
    \frq(Q, G) = \{(u, |\img(u)|)~|~u\in V_p\}, 
\end{equation}
}
\noindent where $\img(u)$ is the image of a pattern node $u$ in $G$. 
\eat{It can be easily verified that this support measure is {\em anti-monotonic}.}

\eat{
\begin{example}
\label{exa-metrics}
Recall graph $G$, pattern $Q_1$ and its matches in Fig.~\ref{fig:example1}. 
It can be easily verified that for $Q_1$, $\img(\kw{C})$=$\{\kw{C_0},\kw{C_1},\kw{C_2}\}$,
$\img(\kw{LB})$=$\{\kw{LB_0}, \kw{LB_1}, \kw{LB_2}\}$,
$\img(\kw{EB})$=$\{\kw{EB_0}, \kw{EB_1}\\, \kw{EB_2}\}$,
$\img(\kw{S})$=$\{\kw{S_0},\kw{S_1},\kw{S_2},\kw{S_3},\kw{S_4},\kw{S_5}\}$,
which leads to $\supp(Q_1,G)=3$ rather than 12. \eop
\end{example}
}

\begin{example}
\label{exa-metrics}
Recall graph $G$, pattern $Q_1$ and its matches in Fig.~\ref{fig:example1}. It is easy to see that
$\img({\rm DBA})$=$\{{\rm v_8}, {\rm v_9}, {\rm v_{19}}, {\rm v_{20}}, {\rm v_{21}}\}$,
$\img({\rm ST})$=$\{{\rm v_{11}}, {\rm v_{13}}, {\rm v_{22}}, {\rm v_{23}}, {\rm v_{24}}\}$,
$\img({\rm PRG})$=$\{{\rm v_{10}},{\rm v_{12}}, {\rm v_{25}}, {\rm v_{26}}, {\rm v_{27}}, {\rm v_{28}}\}$, which leads to $Sup(Q_1,G) = 5$. \looseness=-1
\eat{ 
Recall graph $G$, pattern $Q_1$ and its matches in Fig.~\ref{fig:example1}. It is easy to verify that 
$\img({\rm DBA})$=$\{{\rm v_8}, ...\} = 5$, 
$\img({\rm ST})$=$\{{\rm v_{11}}, ...\} = 5$, 
$\img({\rm PRG})$=$\{{\rm v_{10}}, ...\} = 6$, which leads to $Sup(Q_1,G)=5$. 
}
\eat{
Recall graph $G$, pattern $Q_1$ and its matches in Fig.~\ref{fig:example1}. It is easy to verify that 
$\img({\rm DBA})$=$\{{\rm v_8}, {\rm v_9}, {\rm v_{19}}, ...\} = 5$, 
$\img({\rm ST})$=$\{{\rm v_{11}}, {\rm v_{13}}, {\rm v_{22}}, ...\} = 5$, 
$\img({\rm PRG})$=$\{{\rm v_{10}},{\rm v_{12}}, {\rm v_{25}}, ...\} = 6$, which leads to $Sup(Q_1,G)=5$.  
}
\eop
\end{example}

\begin{table}[t]
\begin{center}
\caption{A summary of notations} \label{tab-terms}
\begin{footnotesize}
\begin{tabular}
{|c|c|}
\hline \textbf{Symbols} & \textbf{Notations} \\
\hline
\hline $G=(V, E, L)$ & a data graph \\
\hline $Q=(V_p,E_p,f_v)$ & a pattern \\
\hline $G_s\subseteq G$ & $G_s$ is a subgraph of $G$ \\
\hline $Q'\sqsubseteq Q$ & $Q'$ is a sub-pattern of $Q$ \\
\hline
\hline $M(Q,G)$ & the set of matches of $Q$ in $G$ \\
\hline $\img(u)$ & the set of matches of node $u$ of $Q$ in $G$, derived from $M(Q,G)$ \\
\hline
\hline $|V|+|E|$ & $|G|$, the size of $G$  \\ 
\hline $|V_p|+|E_p|$ & $|Q|$, the size of $Q$ \\
\hline 
\hline ${\cal T}$ & a rooted and directed tree for maintaining frequent patterns \\
\hline $h$ & the height of tree ${\cal T}$ \\
\hline $\supp(Q,G)$ (resp. $\theta$) & the support of a pattern $Q$ in $G$ (resp. threshold of support)\\
\hline $\textsc{Itrs}(Q)$ & the interestingness of a pattern $Q$ \\
\hline $\widehat{Q}$ & the {\em complete} pattern of $Q$ \\
\hline $D(Q)$ (resp. $D_i(Q)$) & the domain of a pattern $Q$ (resp. a pattern node $u_i$ in $Q$) \\
\hline $e_x=(u_i,u_j)$ & an edge for pattern extension \\
\hline
\end{tabular}
\end{footnotesize}
\end{center}


\end{table}

\begin{definition}
{\bf Frequent Pattern Mining.} Given a graph $G$ and an integer $\theta$ as the support threshold, it is to discover a set $\mathbb{S}$ of frequent patterns $Q$ in $G$ such that $\supp(Q, G)\geq \theta$ for any $Q$ in $\mathbb{S}$. 
\eop
\end{definition}


In practice, the task of FPM faces three challenges: (1) the underlying graphs $G$ are typically very large, and in the meanwhile, the FPM problem is intractable, it is hence very costly to identify all the frequent patterns on such large graphs; (2) there may return excessive patterns which bring trouble to users' inspection and application, moreover people are more interested in those patterns which are top ranked~\cite{YanH03}; and (3) it is not easy to set a viable support threshold $\theta$, because a large (resp. small) $\theta$ will lead to too few (resp. many) patterns~\cite{YanH03}. In light of these, it is necessary to investigate the {\em top-$k$ pattern mining} problem. While, to do this, it is crucial to develop a metric for measuring the interestingness of a pattern. 

Existing metrics for measuring patterns' interestingness can be divided into two types: subjective metrics and objective metrics. A formalization of subjective metric was first introduced by \cite{LeeuwenBSM16}, followed by several similar counterparts. All these metrics, however, are based on information theory and are computationally expensive. In contrast, objective metrics \cite{YanH03,huan2004spin,ChiXYM05} consider the structural containment relationship among patterns, on the basis of the ``closeness'' property, resulting in better efficiency. 
Inspired by the objective metrics, in this paper, we evaluate the interestingness of a pattern $Q = (V_p, E_p, f_v)$ as,

\begin{equation}
    \textsc{Itrs}(Q)=|Q|=|V_p|+|E_p|.
\end{equation}

\begin{example}
Recall patterns $Q_1$, $Q_2$ and $Q_3$ in Fig.~\ref{fig:example1} (b). One may verify that $\textsc{Itrs}(Q_1)=6, \textsc{Itrs}(Q_2)=8$ and $\textsc{Itrs}(Q_3)=10$. Among three patterns, $Q_3$ is considered more interesting, as it subsumes others; in addition, it is frequent entails that the others are frequent as well. \eop 
\end{example}

Indeed, the metric is a simplified closeness-based metric, as it simplifies evaluation of pattern containment with pattern size. Moreover, it is cheaper to evaluate and can be adapted based on practical requirements, e.g., by integrating to developing a {\em top-$k$ pattern mining} algorithm with {\em early termination} property.

\stitle{Problem formulation}. The \textsc{TopkPM} problem is formalized as follows.
\begin{itemize}
    \item Input: A single large graph $G$, support threshold $\theta$ and integer $k$.
    \item Output: A set $\mathbb{S}_k$ of patterns $Q$ discovered from $G$ such that $|\mathbb{S}_k|\leq k$, $\supp(Q, G)\geq \theta$ for any $Q$ in $\mathbb{S}_k$ and $\arg\max_{\mathbb{S}_k\subseteq \mathbb{S}} \sum_{Q\in \mathbb{S}_k} \textsc{Itrs}(Q)$.
\end{itemize}
Intuitively, the problem is to find a set of $k$ (specified by users) patterns that not only satisfy support constraint but also take the largest sum of interestingness values. However, the problem is nontrivial.  

\begin{prop}
\label{prop-1}
The decision problem of \textsc{TopkPM} is NP-hard. 
\end{prop}

To see Prop.~\ref{prop-1}, observe that the subgraph isomorphism (ISO) problem is embedded in \textsc{TopkPM} problem, thus \textsc{TopkPM} problem must be at least as hard as ISO problem. Since ISO is an NP-complete problem~\cite{isomorphism}, thus \textsc{TopkPM} problem must be NP-hard.

\eat{Relatively, the approximation algorithm relaxes a restriction,
\begin{equation}
    \arg\max_{\mathbb{S}_k\subseteq \mathbb{S}} \sum_{Q\in \mathbb{S}_k} \kw{itrs}(Q),
\end{equation}
to return results faster. But the approximate algorithm strives to approach this goal.}


To tackle the issue, one may develop an algorithm (\naive) that applies a ``find-all-select'' strategy to identify top-$k$ patterns. In a nutshell, \naive discovers a complete set $\mathbb{S}$ of frequent patterns by using any existing frequent pattern mining algorithm, ranks frequent patterns according to their interestingness values and picks $k$ best ones. Though straightforward, \naive has to mine all the frequent patterns, hence is prohibitively expensive and even not doable on large graphs. To rectify this, one can incorporate both early termination strategy and approximation scheme. We next illustrate more in Section~\ref{sec-dis}. 



\section{Mining Near-Optimal Top-$k$ Patterns}
\label{sec-dis}

In this section, we first outline an algorithm that preserves early termination property, for identifying near-optimal top-$k$ patterns. We then present a novel method for estimating \mni. 

\eat{
\begin{theorem}
\label{theo-alg}
Given a graph $G$, support threshold $\theta$ and an integer $k$, there exists an algorithm \aprtopk for the \textsc{TopkPM} problem, that finds $k$ patterns such that (a) the computational complexity of \aprtopk is in $O(|V|^{n}\cdot n^{n+1})$ time ($n$ refers to the expansion times), and (b) \aprtopk can terminate as soon as $k$ qualified patterns are discovered. 
\end{theorem}

We next prove Theorem~\ref{theo-alg} by presenting an algorithm as a constructive proof. 
}

\subsection{Mining with Early Termination}
\label{sec-alg}

By Prop.~\ref{prop-1}, we know that identifying the optimal top-$k$ patterns requires extremely high computational costs, which is infeasible in practice. Hence, an algorithm that is able to efficiently discover near-optimal top-$k$ patterns is more desired. 
Motivated by this, we develop such an algorithm, denoted as \aprtopk. 

In contrast to traditional methods, \aprtopk works in an incremental manner to identify top-$k$ patterns, during the period, compact data structures are used for estimating pattern supports. These together significantly lower both computational and space costs while retaining near-optimal recall.


\eat{
\begin{figure}[tb!]
\begin{center}
{\small
\begin{minipage}{4.7 in}
\myhrule
\vspace{-1ex}
\mat{0ex}{
{\bf Algorithm}~\aprtopk~\\
\sstab {\sl Input:\/} \ Graph $G$, support threshold $\theta$, an integer $k$, a random pick parameter $m$.\\
{\sl Output:\/} A set of no more than $k$ patterns. \\
\bcc \hspace{1ex} initialize $flag :=\textbf{false}$; $\mathbb{S}_k:=\varnothing$; $L :=\varnothing$; ${\cal T}$ as an empty tree;\\
\icc \hspace{1ex} initialize $\kw{fEdges}$; update ${\cal T}$;\\
\icc \hspace{1ex} \While($flag \neq \textbf{true}$)~\Do \\
\icc \hspace{3ex}   $L := \forwardtree(\kw{fEdges},{\cal T})$; \\
\icc \hspace{3ex}   \For \Each pattern $Q_c$ in $L$ \Do \\
\icc \hspace{5ex}       Domain $D:= Q_c$.$\kw{Parent}.$Domain;\\
\icc \hspace{5ex}       \If $\te(G,Q_c,D,\theta,m)$ $\geqslant$ $\theta$ \Then update ${\cal T}$ with $Q_c$; \\
\icc \hspace{3ex}   \If ${\cal T}$ was not updated \Then $flag :=\textbf{true}$;\\ 
\icc \hspace{1ex} $\mathbb{S}_k:=\topksch({\cal T}, \theta, k)$; \\
\icc \hspace{0ex} \Return $\mathbb{S}_k$;\\
}

\vspace{-5ex}
\mat{0ex}{
{\bf Procedure}~\topksch \\
{\sl Input:\/} \= Graph $G$, a tree ${\cal T}$, support threshold $\theta$, integer $k$. \\
{\sl Output:\/} A set of patterns. \\
\bcc \hspace{1ex} initialize $Terminate := \textbf{false}$; $\mathbb{S}_k:=\varnothing$; $h$ as the height of ${\cal T}$; \\
\icc \hspace{1ex} \While($Terminate \neq \textbf{true}$) \Do \\
\icc \hspace{3ex} \For \Each $v$ at level $h$ of ${\cal T}$ \Do \\
\icc \hspace{5ex} $L := \backwardtree(Q_{[v]}, \kw{fEdges})$; \\
\icc \hspace{5ex}   \For \Each pattern $Q_c$ in $L$ \Do \\
\icc \hspace{7ex}       \If $\te(G,Q_c,D_s,\theta,m)$ $\geqslant$ $\theta$ \Then $\mathbb{S}_k:=\mathbb{S}_k\bigcup \{Q_c\}$; \\
\icc \hspace{7ex} \If termination condition is satisfied \Then \\ 
\icc \hspace{9ex} $Terminate :=\textbf{true}$; update $\mathbb{S}_k$; \Break; \\
\icc \hspace{3ex} update $h$; \\
\icc \hspace{1ex} \Return $\mathbb{S}_k$; \\
}
\vspace{-6ex}
\myhrule
\end{minipage}
}
\end{center}
\vspace{-2ex}
\caption{Algorithm \aprtopk} \label{alg:ptnmine}
\vspace{-3ex}
\end{figure}
}

\eat{
\begin{algorithm}
    \caption{\aprtopk}
    \label{aprtopk}
    \begin{algorithmic}[1]
        \Require Graph $G$, support threshold $\theta$, an integer $k$, a random pick parameter $m$.
        \Ensure A set of no more than $k$ patterns.
        \State initialize $flag :=\textbf{false}$; $\mathbb{S}_k:=\varnothing$; $L :=\varnothing$; ${\cal T}$ as an empty tree;
        \State initialize $\kw{fEdges}$; update ${\cal T}$;
        \While{$flag \neq \textbf{true}$}
            \State $L := \forwardtree(\kw{fEdges},{\cal T})$; 
            \For{\textbf{each} pattern $Q_c$ in $L$}
                \State Domain $D:= Q_c$.$\kw{Parent}.$Domain;
                \If{$\te(G,Q_c,D,\theta,m)$ $\geqslant$ $\theta$}
                    \State update ${\cal T}$ with $Q_c$; 
                \EndIf
            \EndFor
            \If{${\cal T}$ was not updated}
                \State $flag := \textbf{true}$;
            \EndIf
        \EndWhile
        \State $\mathbb{S}_k:=\topksch({\cal T}, \theta, k)$;
        \State \Return $\mathbb{S}_k$;

    \end{algorithmic}
\end{algorithm}

\begin{algorithm}
    \caption{\topksch}
    \begin{algorithmic}[1]
        \Require Graph $G$, a tree ${\cal T}$, support threshold $\theta$, integer $k$.
        \Ensure A set of patterns.
        \State initialize $Terminate :=\textbf{false}$; $\mathbb{S}_k:=\varnothing$; $h$ as the height of ${\cal T}$;
        \While{$Terminate \neq \textbf{true}$}
            \For{\textbf{each} $v$ at level $h$ of ${\cal T}$}
                \State $L := \backwardtree(Q_{[v]}, \kw{fEdges})$;
                \For{\textbf{each} pattern $Q_c$ in $L$}
                    \If{$\te(G,Q_c,D_s,\theta,m)$ $\geq  \theta$}
                        \State $\mathbb{S}_k:=\mathbb{S}_k\bigcup \{Q_c\}$;
                    \EndIf
                    \If{termination condition is satisfied}
                        \State $Terminate := \textbf{true}$; update $\mathbb{S}_k$; \textbf{Break};
                    \EndIf
                \EndFor
            \EndFor
            \State update $h$;
        \EndWhile
        \State \Return $\mathbb{S}_k$;
    \end{algorithmic}
\end{algorithm}
}

\begin{algorithm}[t]
    \caption{\aprtopk}
    \label{aprtopk}
    \begin{algorithmic}[1]
        \Require Graph $G$, support threshold $\theta$, integers $k$ and $m$.
        \Ensure A set of no more than $k$ patterns.
        \State initialize $flag := \textbf{false}$; $\mathbb{S}_k:=\varnothing$; $L :=\varnothing$; ${\cal T}$ as an empty tree;
        \State initialize $fEdges$; update ${\cal T}$;
        \While{$flag \neq \textbf{true}$}
            \State $L := \forwardtree(fEdges,{\cal T})$; 
            \For{\textbf{each} pattern $Q_c$ {\bf in} $L$ }
                \If{$\te(G,Q_c,D(Q_p),\theta,m)$ $\geqslant$ $\theta$}
                    \State update ${\cal T}$ with $Q_c$; 
                \EndIf
            \EndFor
            \If{${\cal T}$ was not updated}
                \State $flag := \textbf{true}$;
            \EndIf
        \EndWhile
        \State $\mathbb{S}_k:=\topksch({\cal T}, \theta, k, m)$;
        \State \Return $\mathbb{S}_k$;

        \Statex
        \Function{\topksch}{${\cal T}$, $\theta$, $k$, $m$}
            \State initialize $Terminate := \textbf{false}$; $\mathbb{S}_k:=\varnothing$; $L :=\varnothing$; $h$ as the height of ${\cal T}$;
            \While{$Terminate \neq \textbf{true}$}
                \For{\textbf{each} $v$ at level $h$ of ${\cal T}$}
                    \State $L := \backwardtree(Q_{[v]}, fEdges)$;
                    \For{\textbf{each} pattern $Q_c$ {\bf in} $L$}
                        \If{$\te(G,Q_c,D(Q_p),\theta,m)$ $\geq  \theta$}
                            \State $\mathbb{S}_k:=\mathbb{S}_k\bigcup \{Q_c\}$;
                        \EndIf
                        \If{termination condition is satisfied}
                            \State $Terminate :=\textbf{true}$; 
                            \State update $\mathbb{S}_k$;     
                            \State \textbf{break};
                        \EndIf
                    \EndFor
                \EndFor
                \State update $h$;
            \EndWhile
            \State \Return $\mathbb{S}_k$;
        \EndFunction
    \end{algorithmic}
\end{algorithm}

\eat{
(a) incrementally identifies top-$k$ patterns, such that it can terminate earlier; and (b) applies an approximation scheme to enhance the efficiency at the cost of slightly decreased accuracy. 
}

\stitle{Framework}. As shown in the Pseudo-Code in Algorithm \ref{aprtopk}, \aprtopk takes a single (possibly large) graph $G$, a support threshold $\theta$, an integer $k$ and a parameter $m$ as input and returns a set $\mathbb{S}_k$ of qualified patterns that are close to the optimal solution as output. Here, parameter $m$ is used to limit the operation times of procedure \textsc{NodeChoose} (see Eq. 3), thereby improving efficiency. During mining, \aprtopk performs three main tasks, \ie Initialization (lines 1-2), Tree patterns identification (lines 3-9), and Top-$k$ patterns mining (line 10). All the frequent patterns are organized in a directed tree ${\cal T}$, which is dynamically maintained. In particular, the growth of ${\cal T}$ follows a bottom-up manner, starting from ``seed'' patterns (see below for explanations). To simplify discussion, we use ``parent'' (resp. ``child'') to denote relationship of two patterns which correspond to parent-child nodes in ${\cal T}$.

\etitle{Initialization}. Four parameters are initialized, \ie a boolean variable $flag$ to control {\bf while} loop, an empty set $\mathbb{S}_k$ for keeping track of top-$k$ patterns, an empty set $L$ for maintaining candidate patterns and an empty tree ${\cal T}$ to record frequent patterns (line 1). Later on, frequent single-edge patterns (\aka ``seed patterns'') are identified. They are included in a set $fEdges$ and used to update ${\cal T}$ (line 2). Note that, after initialization, tree ${\cal T}$ is consisted of isolated nodes that correspond to ``seed patterns'' in $fEdges$. \looseness=-1

\eat{
\aprtopk then identifies frequent single-edge patterns (\aka ``seed patterns'') which are used to initialize $fEdges$ and update ${\cal T}$ (line 2). After initialization, ${\cal T}$ consists of a set of nodes that correspond to ``seed patterns'' in $fEdges$. 
}


\eat{
\begin{figure}
\centering
\includegraphics[width=3.33in, keepaspectratio]{fig/pattern.eps}
\caption{the main components of pattern}
\label{fig:pattern}
\end{figure}
}

\etitle{Tree patterns identification}. In this stage, \aprtopk iteratively identifies frequent ``tree'' patterns, following a {\em level-wise} strategy (lines 3-9).
In each round iteration, \aprtopk performs as follows. (1) It generates a set $L$ of ``tree'' patterns as candidates with procedure~\forwardtree (line 4). 
Note that \forwardtree (not shown) produces candidate patterns by expanding ``tree'' patterns that locate at the top level of ${\cal T}$ with ``seed patterns'', following forward expansion (See Def.~\ref{def-expansion}). (2) For each candidate pattern $Q_c$, \aprtopk employs a procedure \te to calculate its support and updates ${\cal T}$ with $Q_c$ if it is frequent (lines 6-7). The details of \te for support estimation will be given shortly. (3) After above process, if ${\cal T}$ remains unchanged, the flag $flag$ is then updated as \textbf{true}, indicating that no new pattern was generated and the {\bf while} loop no longer needs to continue (lines 8-9). By now, ${\cal T}$ grows into a tree with nodes corresponding to frequent tree patterns, and will be used for further processing. \looseness=-1

\begin{figure}[t]
\includegraphics[width=\linewidth]{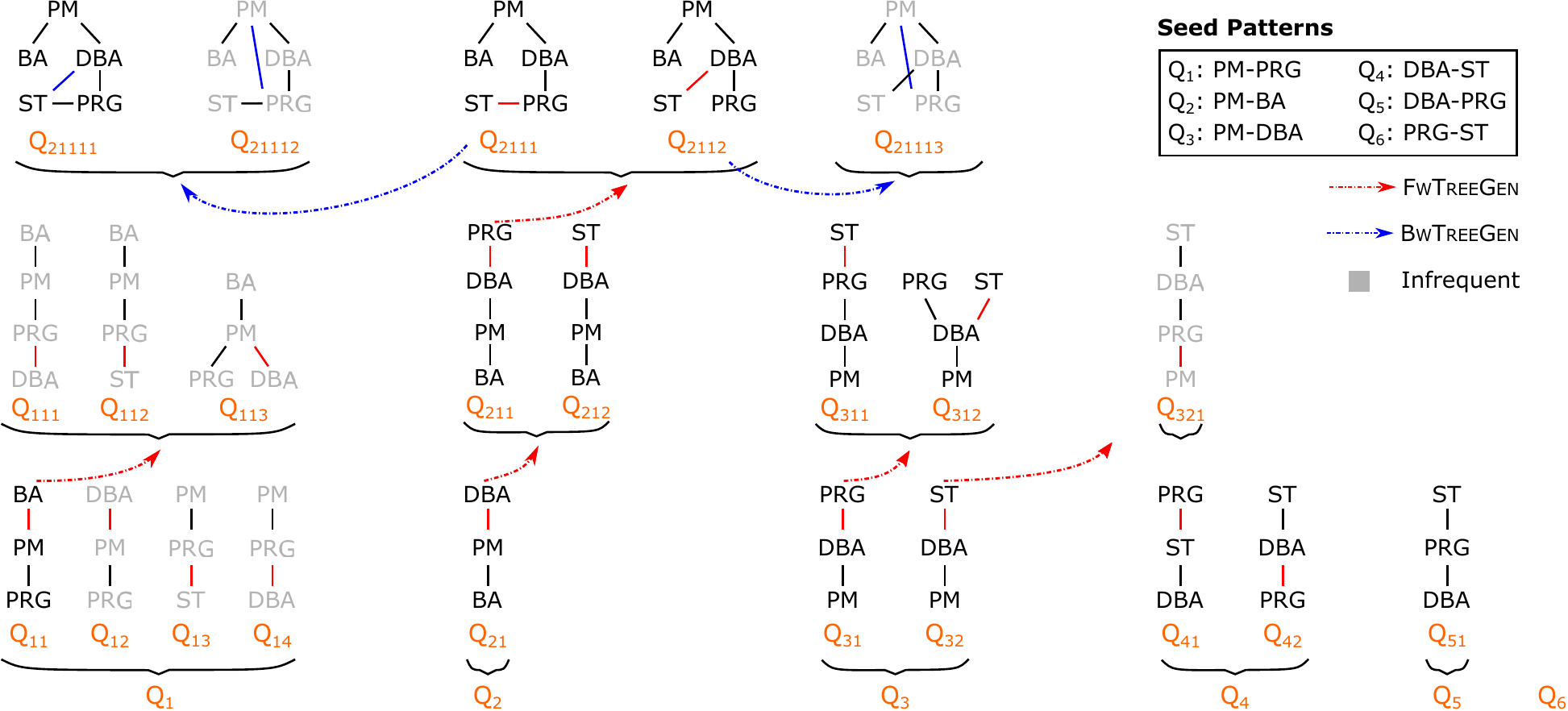}
\centering
\caption{Growth of ${\cal T}$, via {\em forward} and {\em backward} expansions. Infrequent patterns are marked in grey.}
\label{fig-running-example}
\end{figure}

\begin{example}
\label{exa-alg-tree}
On graph $G$ of Fig.~\ref{fig:example1} (a), \aprtopk first identifies frequent edges as seed patterns $Q_1$-$Q_6$ (shown in Fig.~\ref{fig-running-example}), as their supports are no less than 3. Then, \aprtopk applies \forwardtree to generate candidate patterns following {\em forward expansion}, in a {\em level-by-level} manner. For example, using pattern $Q_1$, \aprtopk generates candidate patterns by enlarging $Q_1$ with other frequent ``seed'' patterns and produces $L =\{Q_{11}, Q_{12}, Q_{13}, Q_{14}\}$. As patterns in $L$ are generated via {\em forward expansion}, their forward edges in Fig.~\ref{fig-running-example} are marked in red. Four levels of ``nontrivial'' candidate patterns (patterns without duplicate node labels) are listed in Fig.~\ref{fig-running-example}. \looseness=-1
\eop
\end{example}

\etitle{Top-$k$ patterns mining}. Based on frequent tree patterns, the procedure~\topksch is employed to discover top-$k$ patterns. Specifically, \topksch first initializes necessary parameters: a Boolean variable $Terminate$ as a flag for loop control, an empty set $\mathbb{S}_k$ to store $k$ chosen patterns and an integer $h$ as the height of ${\cal T}$ (line 13). \topksch then iteratively generates non-tree patterns and identifies top-$k$ ones, where the pattern generation process starts from tree-patterns located at the top level of ${\cal T}$, and follows a top-down manner (lines 14-24). In each round iteration, \topksch selects a node $v$ (corresponding to a ``tree'' pattern $Q_{[v]}$) at level $h$ of ${\cal T}$, and generates a set $L$ of candidate patterns with \backwardtree (line 16). Note that \backwardtree (not shown) works along the same line as \forwardtree, but only enlarges a pattern $Q_{[v]}$ with ``seed patterns'' via {\em backward expansion} (See Def.~\ref{def-expansion}). For each candidate pattern $Q_c$ in $L$, \topksch verifies its support still with \te and enlarges $\mathbb{S}_k$ with $Q_c$ if it is a qualified pattern (line 19). \topksch next verifies whether the termination condition, specified by Proposition~\ref{prop-early-termination}, is satisfied (line 20). \looseness=-1

\eat{
\begin{example}
\label{exa-alg-tree}
On graph $G$ of Fig.~\ref{fig-Auxiliary} (a), We suppose support threshold $\theta = 3$. \aprtopk first finds out the frequent edges and converts them into edge patterns $Q_1$-$Q_5$, as their supports all equal to $3$. Then, \aprtopk applies \kw{ForwardTree}(resp. \kw{BackwardTree}) to generate candidate patterns, in a {\em level-by-level} manner. For example, using pattern $Q_1$, \aprtopk generates candidate patterns by enlarging $Q_1$ with other frequent single-edge patterns and produces $L =\{Q_{11},Q_{12},Q_{13},Q_{14},Q_{15}\}$. Four levels candidate patterns are shown in Fig.~\ref{fig-running-example} (b), where the red arrow represents the new forward edge generated during the process of \kw{ForwardTree}, and the blue arrow represents the new backward edge generated during the process of \kw{BackwardTree}. 
\end{example}
}

\begin{prop}
\label{prop-early-termination}
Given parameters $\theta$, $k$ and a tree ${\cal T}$, whose nodes correspond to the set $\mathbb{S}^t$ of frequent ``tree'' patterns, a $k$-element set $\mathbb{S}_k$ is the top-$k$ pattern set, if (1) $\supp(Q, G)\geq \theta$ for each $Q$ in $\mathbb{S}_k$, and (2) $\min\{\textsc{Itrs}(Q)|Q\in \mathbb{S}_k\}\geq\max\{\textsc{Itrs}(\widehat{Q_t})|Q_t\in \overline{\mathbb{S}^t})$. 
\end{prop}

Here, $\overline{\mathbb{S}^t}$ is a subset of $\mathbb{S}^t$ and includes those tree patterns that have not been used for pattern expansion, and $\widehat{Q_t}$ is a {\em complete} pattern of a tree pattern $Q_t$ in $\overline{\mathbb{S}^t}$. Observe that $\textsc{Itrs}(\widehat{Q_t})$ must be larger than interestingness value of any other pattern that is expanded from $Q_t$, as a result, when the minimum interestingness value of a pattern $Q$ in $\mathbb{S}_k$ is already no less than the maximum interestingness value of the complete pattern of a tree pattern $Q_t$ in $\overline{\mathbb{S}^t}$, then $\sum_{Q\in \mathbb{S}_k} \textsc{Itrs}(Q)$ is already maximized and no further exploration is needed. 

Indeed, Prop.~\ref{prop-early-termination} enables algorithm \aprtopk to terminate earlier. As top-$k$ patterns mining are performed in a top-down manner, the set $\overline{\mathbb{S}^t}$ (initially the same as $\mathbb{S}^t$) is hence dynamically updated with a seen $Q_t$. \looseness=-1

\vspace{0.5ex}
If the termination condition is satisfied, \topksch sets $Terminate$ as \textbf{true}, eliminates redundant patterns in $\mathbb{S}_k$ if $|\mathbb{S}_k|>k$, breaks the {\bf while} loop (lines 21-23) and returns $\mathbb{S}_k$ as final result (line 25). Otherwise, \topksch decreases $h$ by $1$ (line 24), indicating that a new round selection will start from level $h$-1 of ${\cal T}$.  \looseness=-1

\begin{example}
\label{exa-topk-search}
Recall Example~\ref{exa-alg-tree}. To mine the top-1 pattern on graph $G$ of Fig.~\ref{fig:example1} (a), a pattern $Q_{21111}$ is generated via backward expansion (marked in blue line for backward edges) from its parent $Q_{2111}$ at level $4$. 
\topksch then evaluates its support and enlarges $\mathbb{S}_k$ with it. The above process terminates until candidates generated from patterns at level $4$ of ${\cal T}$ (Fig.~\ref{fig-running-example}) are all processed, as the remaining candidates can not have higher \textsc{Itrs} values. \looseness=-1 \eop
\end{example}

\eat{
\stitle{forWardExtend}. forWardExtend and backwardExtend are the core parts of our algorithm. Here, we will introduce to you in detail.
Every time the pattern is extended, the pattern will record its own expansion information and store it in a list $infoList$. Through this list, algorithm can use the \mni table of the pattern to recursively generate the corresponding embeddings, and use it to calculate the \mni of the new pattern. In procedure \textbf{forWardExtend}, The algorithm will extend the pattern $Q_c$ for each pattern that meets the threshold in the previous iteration, and calculate the \mni support for it(lines1-5). More details, in procedure \textbf{calMNI}, We first initialize a stack to store the vertices of recursive traversal. Then, the algorithm begins to recursively traverse the \mni table of the new 
pattern to the parent pattern. At each level of recursion, the point in the next column in the \mni table will be taken out and the neighbor vertices will be intersected(lines 11-13). Traverse this vertices set and verify that the vertex meets the requirements before adding it to the stack(lines 15-20). When a certain recursion, the added point makes the size of the stack reach the number of vertices of the parent mode graph, we think that this is a match corresponding to the parent mode, the algorithm uses this match to expand, and the corresponding point is added to the \mni of the new mode Table.
}

\eat{
\etitle{Correctness}. The correctness of \ptnmine is warranted by the following observations. (1) The pattern generation scheme will never miss any qualified pattern. (2) The support computation with {\em partial evaluation} is correct. (3) The strategy used by top-$k$ pattern selection will never choose a pattern whose interestingness is less than any top-$k$ pattern. \looseness=-1
}

\eat{
Firstly, \ptnmine employs two procedures \kw{TreeGen} and \kw{NonTreeGen} to iteratively generate candidate patterns. For \kw{TreeGen}, it enlarges an existing frequent ``tree'' pattern with a frequent single-edge pattern, by following {\em forward expansion}. It can be verified via induction that for all frequent patterns in a graph $G$, each of them must have a spanning tree, that is generated by \kw{TreeGen}. Based on the ``tree'' pattern set, \kw{NonTreeGen} generates candidate patterns via {\em backward expansion}. One may also verify that no frequent pattern will be missed. 

Secondly, \ptnmine computes the support of a candidate pattern $Q_c$ in a decompose-and-assemble way. Local matches and their support of $Q_c$ can be easily and correctly identified. To verify the existence of a match of $Q_c$ that cross multiple sites, \ptnmine collects all the ``virtual matches'' which overlap with each other and then solves a satisfiability problem. The support of $Q_c$ can then be correctly obtained by adding them together. 

Thirdly, note that the top-$k$ patterns are generated based on a set of frequent ``tree'' patterns, in addition, the $\kw{itrs}$ value of a ``tree'' pattern $Q_t$ can not exceed its corresponding complete graph $\widehat{Q_t}$. Thus, when the possibly largest $\kw{itrs}$ value is already no larger than the smallest $\kw{itrs}$ value of $\mathbb{S}_k$, then $\mathbb{S}_k$ must be the top-$k$ pattern set and the algorithm can terminate immediately. 

Putting (1),(2) and (3) together, the correctness of \ptnmine follows.  
}

\eat{
\etitle{Complexity}. For a pattern $Q_c$ with vertex set $V_{p_c}$, there may exist at most $|V|^{|V_{p_c}|}$ matches in a graph $G$. Thus, it takes $O(|V|^{|V_{p_c}|+1})$ time to identify matches of a candidate pattern $Q_{c}'$ ($Q_c\sqsubseteq Q_c'$). If $Q_c$ is a ``tree'' pattern, there may generate at most $|V_{p_c}||{\cal L}|$ candidate patterns after {\em forward expansion}; otherwise, at most $O(|V_{p_c}|^2)$ candidate patterns will be generated through {\em backward expansion}. In the meanwhile, it still needs $O(|V|^{|V_{p_c}|+1}\cdot(|V_{p_c}|+1)^{|V_{p_c}|+1})$ time to verify whether $Q_c$ has been generated before, since at most $|V|^{|V_{p_c}|+1}$ candidate patterns may be generated before and each round isomorphism checking needs $O((|V_{p_c}|+1)^{|V_{p_c}|+1})$ time. 
By induction, it takes \ptnmine $\sum_{i\in [1,n]} (|V_{p_c}|+i-1)|V|\cdot |V|^{|V_{p_c}|+i}(1+(|V_{p_c}|+i)^{|V_{p_c}|+i})$ times to verify all the subsequent patterns for a candidate $Q_c$, where $n$ refers to the expansion times and is bounded by $|V|$, $|{\cal L}|$ and $|V_{p_c}|$ are bounded by $|V|$. As the iteration starts from single-edge patterns $Q_c$ and at most $|V|^2$ different $Q_c$ exists, 
hence \ptnmine is bounded by $O(|V|^{n}\cdot n^{n+1})$ time.


The analysis above completes the proof of Theorem~\ref{theo-alg}. \eop
}

\eat{introduce our method, \ie ~\kw{TravelE}, in which the \mni is calculated without storing matches. Then we extend this method to the problem of finding an approximate solution, in which an {\em early termination} property is used to discover top-$k$ patterns.}

\subsection{Supports Evaluation}
\label{sec-te}

During mining, supports evaluation brings two challenges: (a) high computational cost, since it involves expensive isomorphism checking, which may even be performed exponentially many times; and (b) high space cost for recording all the matches and then calculating \mni for each pattern. On large graphs, such high costs are often not affordable. This calls for effective methods to estimate pattern supports efficiently and accurately. 

To this end, we introduce a novel method \te, which incorporates an approximation scheme for support estimation. The Pseudo-Code of \te is shown in Algorithm \ref{alg:fre}. We next introduce its details, starting from the auxiliary structures it uses. \looseness=-1


\begin{figure}[t]
\includegraphics[width=0.9\textwidth]{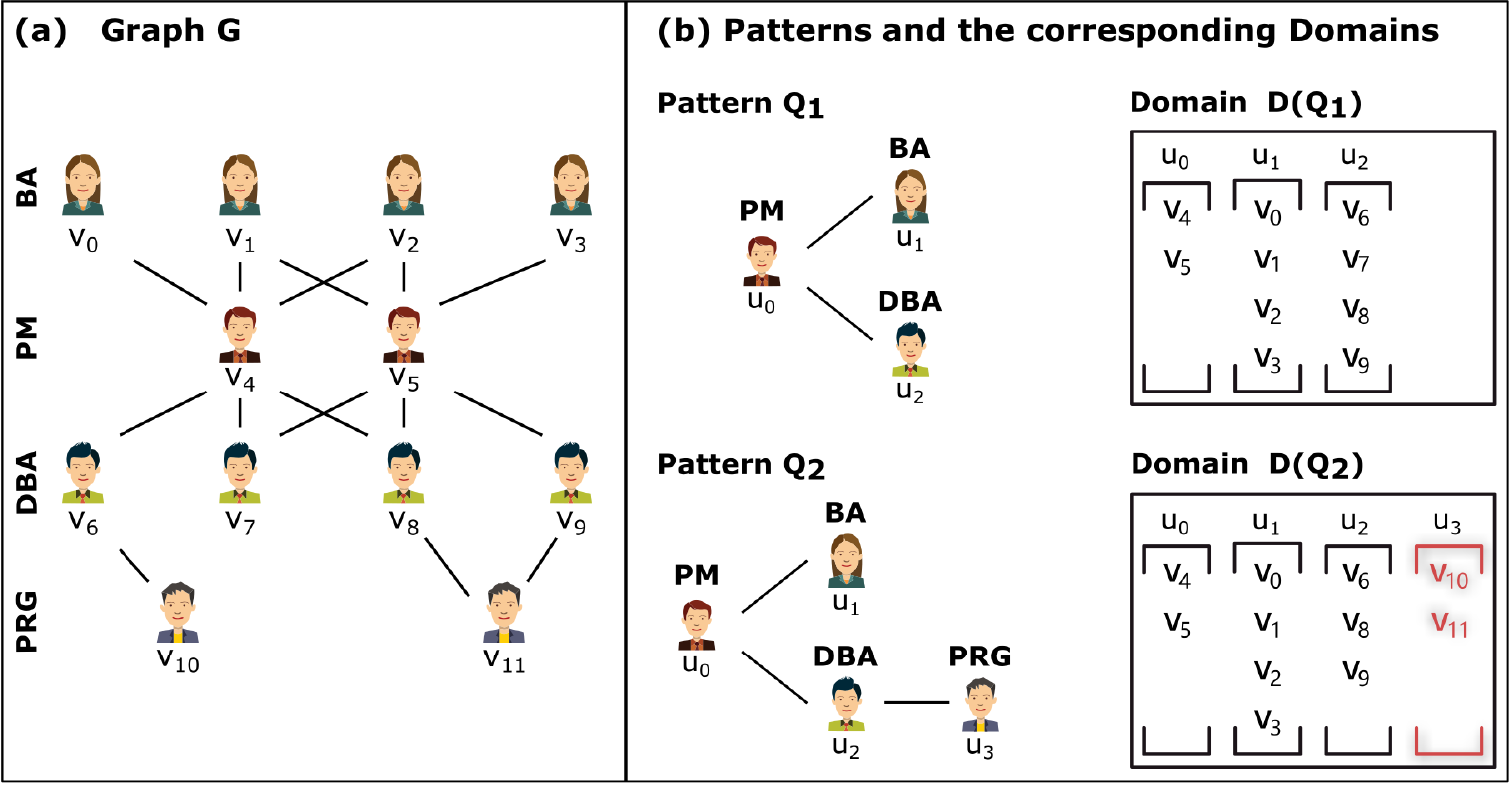}
\centering
\caption{A sample graph, typical patterns and their domains}
\vspace{-2ex}
\label{fig-Auxiliary}
\end{figure}

\eat{
\etitle{\kw{InfoList}}. To avoid costly isomorphism detection, \te records the structural information of a pattern in \kw{InfoList}. That is, for a pattern $Q$, its \kw{InfoList} ${\cal I}$ records how it was generated, \ie the sequence of edge expansion. As will be seen, the structural information will be used to guide the match verification. Note that the \kw{InfoList} ${\cal I}$ of a pattern $Q$ is generated during expansion and updated by either \forwardtree or \backwardtree. 

\begin{example}
\label{exa-alg-InfoList}
Two patterns $Q_1, Q_2$ are shown in Fig.~\ref{fig-Auxiliary} (b), where pattern $Q_1$ is parent pattern of pattern $Q_2$.
At the top is the pattern graph of the patterns.
The pattern graph is obtained by step-by-step extending of the frequent edge pattern.
\kw{InfoList} records how the pattern is expanded step by step.
According to whether the newly added edge is forward or backward, $info$ will be added to Forward or Backward in \kw{InfoList}.
A $info$ \textbf{“0 in 1 C LB"} indicates that node $v_0$ labeled with “C" is pointed by node $v_1$ labeled with “LB" in pattern graph.
When pattern $Q_1$ continues to extend to get pattern $Q_2$, \kw{InfoList} will add one more $info$ \textbf{``1 in 3 LB S''} into Backward.
\end{example}
}

\etitle{Auxiliary Structures.} To facilitate the calculation of $\textsc{MnI}$-based support, an auxiliary structure, called Domain, is used to keep track of matches of a pattern. 

\begin{definition}
{\bf Domain.} Given a graph $G$ and a pattern $Q$ with node set $V_p$, the Domain of $Q$, denoted by $D(Q)$, reorganizes all the matches $M(Q,G)$ of $Q$ in $G$ with a table, whose column head and body correspond to a pattern node $u_i$ ($u_i\in V_p$) and its image $\img(u_i)$, respectively. 
\eop
\end{definition}

Abusing the notation of domain, we use $D_i(Q)$ to indicate the $i$-th domain of $D(Q)$, which essentially corresponds to $\img(u_i)$. 

\begin{example}
\label{exa-alg-Domains}

As shown in Fig.~\ref{fig-Auxiliary}, the match set $M(Q_1,G)=\{(v_0,v_4,v_6),(v_0,v_4,\\v_7),(v_0,v_4,v_8),(v_1,v_4,v_6),(v_1,v_4,v_7),(v_1,v_4,v_8),(v_2,v_4,v_6),(v_2,v_4,v_7),(v_2,v_4,\\v_8),(v_1,v_5,v_7),(v_1,v_5,v_8),(v_1,v_5,v_9),(v_2,v_5,v_7),(v_2,v_5,v_8),(v_2,v_5,v_9),(v_3,v_5,\\v_7),(v_3,v_5,v_8),(v_3,v_5,v_9)\}$ includes in total 18 distinct matches of $Q$ in $G$; while the domain $D(Q_1)$ of $Q_1$ in $G$, shown in Fig.~\ref{fig-Auxiliary}(b) is a more compact data structure, compared with $M(Q_1,G)$. \looseness=-1
\eop
\end{example}


Obviously, the domain of a pattern $Q$ in a graph $G$ is of linear size of $|G|$ and $|Q|$, which is much smaller than $M(Q,G)$ (potentially in exponential size of $|G|$). Apart from compact structure, domains can be used for support estimation efficiently and accurately. \looseness=-1



\eat{
\begin{figure}[tb!]
\begin{center}
{\small
\begin{minipage}{4.5in}
\myhrule
\vspace{-1ex}
\vspace{-4.5ex}
\mat{0ex}{
{\bf Procedure}~Origin \te~\\
{\sl Input:\/} A pattern $Q_c$, Domain $D_s$ of $Q_c$, and parameters $\theta$, $M$ ($M\in (0,1)$).\\
{\sl Output:\/} The \mni of pattern $Q_c$. \\
\bcc \hspace{1ex} initialize a stack $S_d:=\varnothing$ for storing matches;  \\
\icc \hspace{1ex} \For each \kw{AS} in Domain \Do; \\
\icc \hspace{3ex}   start \kw{DFS} from head fraction and constantly execute \kw{DFS}: \\
\icc \hspace{5ex}       Set nodes found acording to \kw{InfoList} as candidate traversal nodes; \\
\icc \hspace{5ex}       $S_d$ push the node that \kw{DFS} visited in current Iteration; \\
\icc \hspace{5ex}           \If the nodes in stack $S_d$ can form a match of pattern $Q_c$ \Then \\
\icc \hspace{7ex}          put the nodes into the corresponding columns of $Q_c$.Domain; \\
\icc \hspace{3ex}  $size_1:=$ size of $v_0$ Domain of $Q_c$; \\
\icc \hspace{3ex}  $size_2:=$ size of untraveled AS; \\
\icc \hspace{2ex}  \If  $size_1 + size_2 < \theta * M$  \Then \Break; \\
\icc \hspace{0ex} \Return \mni ($Q_c$); \\
}

\vspace{-5ex}
\myhrule
\end{minipage}
}
\end{center}
\vspace{-2ex}
\caption{Algorithm \kw{Origin} \te} \label{alg:originTE}
\vspace{-3ex}
\end{figure}
}

\eat{
\begin{figure}[tb!]
\begin{center}
{\small
\begin{minipage}{4.5in}
\myhrule
\vspace{-2ex}
\mat{0ex}{
{\bf Procedure}~ \te~\\
{\sl Input:\/} Graph $G$, a pattern $Q_c$ {\color{blue} and its \kw{InfoList} ${\cal I}_c$}, a Domain $D$ and a threshold $\theta$ \\
{\sl\/} a random pick parameter $m$. \\
{\sl Output:\/} The \mni support of pattern $Q_c$. \\
\bcc \hspace{1ex} \For each node $v$ in $D[0]$ \Do \\
\icc \hspace{3ex}   $D(Q_c):=\textsc{Traverse}$($Q_c$, ${\cal I}_c$, $D$, $v$); \\
\icc \hspace{3ex}  $size:=|D(Q_c)[0]|$ + $|D[0]|$; \\
\icc \hspace{3ex}  \If $size< \theta$  \Then \Break; \\
\icc \hspace{1ex}  calculate $\supp(Q_c)$ with $D(Q_c)$; \\
\icc \hspace{1ex} \Return $\supp(Q_c)$; \\
}

\vspace{-4.5ex}
\mat{0ex}{
{\bf Procedure}~\textsc{Traverse}\\
{\sl Input:\/} Graph $G$, a pattern $Q_c$ {\color{blue} and its \kw{InfoList} ${\cal I}_c$}, a Domain $D$ and a node $v$.\\
{\sl Output:\/} A Domain $D(Q_c)$. \\
\bcc \hspace{1ex} initialize Domain $D(Q_c):=\varnothing$, stacks $\kw{S}_t:=\varnothing$, $S_d:=\varnothing$;  \\
\icc \hspace{1ex} $\kw{S}_t.push(u_0,v)$; \\
\icc \hspace{1ex} \While($\kw{S}_t\neq \varnothing$): \\
\icc \hspace{3ex}   $\langle u',v'\rangle$ := $\kw{S}_t.pop()$; \\
\icc \hspace{3ex}   $S_d.push(v')$; \\
\icc \hspace{3ex}   \If $|S_d|==|D|$ \Then \\
\icc \hspace{5ex}   $\textsc{Expand}(Q_c,S_d,D(Q_c),G)$; \\
\icc \hspace{5ex}      $S_d.pop()$; \\ 
\icc \hspace{3ex}   \Else \\
\eat{
\icc \hspace{4ex}   $info$ := \kw{InfoList}.get($S_d$.size()); \\
\icc \hspace{4ex}   $nodes$ := get constraint nodes according to $info$ (node consistency and arc consistency);\\}
\icc \hspace{4ex}   $H_c:=\textsc{ConsExtr}({\cal I}_c,u',v',S_d,m)$;\\
\icc \hspace{4ex}   \If $H_c==\varnothing$ \Then $S_d.pop()$; \\
\icc \hspace{4ex}   \Else $\kw{S}_t.push(H_c)$; \\
\icc \hspace{0ex} \Return $D(Q_c)$; \\
}
\vspace{-5ex}
\myhrule
\end{minipage}
}
\end{center}
\vspace{-2ex}
\caption{Algorithm \te} \label{alg:originTE}
\vspace{-3ex}
\end{figure}
}

\begin{algorithm}[t]
    \caption{\te}
    \label{alg:fre}
    \begin{algorithmic}[1] 
        \Require Graph $G$, a pattern $Q_c$, the domain $D(Q_p)$ of $Q_p$, parameters $\theta$ and $m$.
        \Ensure The minimum-image-based support \mni of pattern $Q_c$.
        \State initialize domain $D(Q_c)$, a stack $S_d:=\varnothing$, an integer $counter:=|D_0(Q_p)|$;
        \State update $D(Q_p)$;
            \For{{\bf each} node $v$ {\bf in} $D_0(Q_p)$}
                \State restore $S_d$; $counter:=counter-1$; 
                \State $D(Q_c):=\textsc{Traverse}$($G,Q_c,D(Q_c),D(Q_p),S_d$);
                \If{$|D_0(Q_c)| + counter < \theta$}  \State {\bf break}; \EndIf
            \EndFor
        \State calculate $\supp(Q_c)$ with $D(Q_c)$;
        \State \Return $\supp(Q_c)$;

        \Statex
        \Function {Traverse}{$G,Q_c,D(Q_c),D(Q_p),S_d$}
            \State $H_c:=\textsc{ConsExtr}(G, Q_c, D(Q_p), S_d)$; $c := 0$;
            \State $v :=$ \textsc{NodeChoose($H_c, D(Q_c), S_d, c$)};
            \While{ $v \neq null$}
                \State update $c$;
                \State $S_d.push(v)$;
                \If {$|S_d|==|V_{p_p}|$}
                    \State $\textsc{Expand}(G, Q_c, D(Q_c), S_d)$;
                \Else
                    \State $D(Q_c):=\textsc{Traverse}$($G,Q_c,D(Q_c),D(Q_p),S_d$);
                \EndIf
                \State $S_d.pop()$;
                \State $v :=$ \textsc{NodeChoose($H_c, D(Q_c), S_d, c$)};
            \EndWhile
            \State \Return $D(Q_c)$;
        \EndFunction
        
    \end{algorithmic}
\end{algorithm}

\stitle{Support Estimation}. The support estimation is fulfilled by the procedure \te, which leverages a recursive function \textsc{Traverse} to update domains of candidate patterns. The Pseudo-Code of \te is shown in Algorithm \ref{alg:fre}. As can be seen, the input of \te includes a graph $G$, a candidate pattern $Q_c$, 
whose support needs to be verified, a pattern $Q_p$ with node set $V_{p_p}$ along with its domain $D(Q_p)$, a support threshold $\theta$ and an integer $m$. Here $Q_c$ is deemed as the ``child'' of $Q_p$, as the corresponding node of $Q_c$ on ${\cal T}$ is a child of that of $Q_p$. Indeed, $Q_c$ is extended with an edge $e_x=(u_i, u_j)$ from $Q_p$. If the expansion is a backward expansion, then $u_j$ is already in $Q_c$, otherwise, $u_j$ is a newly introduced node. 
As will be seen, the parameter $m$ is involved for controlling candidate selection. \looseness=-1



First of all, \te initializes an empty domain for $Q_c$, an empty stack $S_d$ for loop and an integer $counter$ as $|D_0(Q_p)|$ for fast verification (line 1). 
As $Q_c$ is extended from its {\em parent} $Q_p$ with the edge $e_x=(u_i,u_j)$, \te utilizes this property to conduct a preliminary pruning by referencing $e_x$, $D(Q_p)$ and $G$ (line 2). Specifically, \te checks each node $v_k$ in $D_i(Q_p)$ and see whether there exists an edge $(v_k,v_k')$ in $G$, where $v_k$ is a {\em match} of $u_i$ of $Q_p$ and $v_k'\sim u_j$. If $v_k$ does not have such a neighbor $v_k'$, then $v_k$ can not be a {\em match} of $u_i$ of $Q_c$ and is marked with a special symbol indicating its invalidity. \looseness=-1

\begin{example}
\label{alg-running-1}
Taking $G$, $Q_1$ as well as its domain $D(Q_1)$ given in Fig.~\ref{fig-Auxiliary} as input, \te first checks whether each node in $D_2(Q_1)$ has a neighbor $v$ such that $v\sim u_3$, as $Q_2$ is expanded with an edge $e_x=(u_2, u_3)$ from $Q_1$. Then, node $v_7$ is identified and marked as invalid, since it has no neighbor labeled as PRG. \looseness=-1
\eop
\end{example}

$\te$ next iteratively updates $D(Q_c)$ via guided traversal from each $v$ in $D_0(Q_p)$ (lines 3-7). During the iteration, \te restores the stack $S_d$ by pushing $v$ on top of it after cleaning, in addition, \te also decreases the $counter$ by 1, indicating that $v$ has been used for verification (line 4). Afterwards, \te calls \textsc{Traverse} to identify qualified matches of $Q_c$ (line 5, details of \textsc{Traverse} will be elaborated shortly). After the traverse, an updated domain $D(Q_c)$ is returned, \te then verifies the satisfiability of a simple rule, i.e., $|D_0(Q_c)|+counter<\theta$. Intuitively, the rule states that if the total number of qualified matches of $u_0$ of $Q_c$ and unverified matches of $u_0$ of $Q_p$ is already less than $\theta$, then $\supp(Q_c)$ must be less than $\theta$ (property of \mni). If the rule is satisfied, \te breaks the loop immediately, since further verification is no longer needed (line 7). When all the candidates of $u_0$ are verified, \te calculates $\supp(Q_c)$ by using $D(Q_c)$ (line 8) and returns it as final result (line 9).  \looseness=-1

\eat{
{\color{red}Firstly, \te checks new constraints of the child pattern relative to the parent pattern, and prunes the nodes without new constraints in advance according to one-hop neighbors (line 1).} For each node $v$ in $D_0(Q_p)$, 
\te repeatedly updates the domain $D(Q_p)$ with procedure \textsc{Traverse} (lines 2-7). During the iteration, \te employs a rule (line 7) to prune infrequent patterns. The rule is stated as follows: if the sum of $|D_0(Q_c)|$ and $|D_0(Q_p)|$ is less than $\theta$, then pattern $Q_p$ can not be a frequent pattern. After loop, \te simply computes $\supp(Q_c)$ with $D(Q_c)$ (line 8) and returns $\supp(Q_c)$ as final result (line 9). We next illustrate \textsc{Traverse} with details. 
}

\etitle{Procedure \textsc{Traverse}}. Recall that the {\bf M}i{\bf n}imum-{\bf I}mage-based {\bf S}upport only concerns the image of each distinguished node of a candidate pattern $Q_c$ rather than the total number of matches of $Q_c$. Thus, we do not need to enumerate all the matches, but try to obtain a domain of $Q_c$, which is as accurate as possible. Based on this observation, our procedure applies a wise strategy to guide search accurately and economically. 
We next present details of \textsc{Traverse}.

Given a stack $S_d$ that contains match candidates, \textsc{Traverse} works as follows. \looseness=-1 

\vspace{0.5ex}
\noindent $\mathbf{Stage (I)}$:  Based on current status (determined by $S_d$), \textsc{Traverse} identifies a set of nodes $H_c$ for further exploration with a procedure \textsc{ConsExtr} (line 11). To do this, \textsc{ConsExtr} (not shown) first identifies an {\em unvisited} edge $e_u=(u_i,u_j)$ of $Q_c$ for guiding next step exploration. The identification of $e_u$ is based on $v_i$, which locates at $S_d$ and exists in $D_i(Q_c)$ (the corresponding domain of $u_i$). 
\textsc{ConsExtr} next collects those nodes $v_j$ in $G$ (resp. $D_j(Q_p)$), that are neighbors of $v_i$ and satisfy $v_j\sim u_j$ if $e_u$ is a forward (resp. backward) edge. Note that the nodes in $D(Q_p)$ that are marked with invalid symbols will be omitted. For each seen edge $e_u$, it is then marked as {\em visited} to avoid repeated visit. 
In addition, \textsc{Traverse} also initializes a parameter $c$ as 0 for controlling node selection. 

\eat{
\noindent (I) It identifies a set $H_c$ of nodes as candidate matches of a pattern node $u_{i+1}$, which is an endpoint of the a pattern edge $(u_i,u_{i+1})$ (line 11). This task is carried out by a procedure \textsc{ConsExtr}. Indeed,  \textsc{ConsExtr} (not shown) collects those nodes $v_k'$ in $G$, that are neighbors of $v_k$ (as matches of $u_i$). It also initializes a parameter $c$ as 0 for node selection. 
}

\vspace{0.5ex}
\noindent $\mathbf{Stage (II)}$: \textsc{Traverse} picks a node from $H_c$, with procedure \textsc{NodeChoose} (not shown) by using below selection criteria (line 12). 
\begin{equation}
            v := \left\{
            \begin{array}{lr}
                v_h \in H_c\setminus D(Q_c),  &  H_c\setminus D(Q_c)\neq \varnothing ~~~(A)\\
                v_h \in H_c\cap D(Q_c), & \ \ H_c\setminus D(Q_c)= \varnothing \land S_d \not\subseteq D(Q_c) \land c<m ~~~(B)\\
                null, & {\rm otherwise} ~~~(C)
            \end{array}
            \right.
\end{equation}
        
Intuitively, Condition A states that \textsc{NodeChoose} prefers to pick a node $v$ that is not in $D(Q_c)$. The reason for the preference lies in that a node that is not in $D(Q_c)$ is beneficial to enlarge $\supp(Q_c)$, since an unvisited node will lead the  traversal to a large part of unvisited area with higher possibility. If $H_c$ is already contained by $D(Q_c)$ (i.e., $H_c\setminus D(Q_c)= \varnothing$), \textsc{NodeChoose} selects a node from $H_c\cap D(Q_c)$ if Condition B is satisfied. 
Here, Condition B enforces extra two restrictions i.e., $S_d \not\subseteq D(Q_c)$ and $c<m$. For the first restriction, it requires that $S_d$ should contain nodes that are not in $D(Q_c)$, since otherwise, current traversal will not bring any new element to  $D(Q_c)$. The second restriction imposes a number constraint, that is \textsc{Traverse} only picks no more than $m$ nodes from $H_c\cap D(Q_c)$. The number of selected nodes is recorded by a parameter $c$, which is updated when $v$ is used for next round traversal (line 14). By introducing an adjustable parameter $m$, the exploration area at current iteration is restricted, thereby reducing the computational costs. 
When both of two conditions can not be satisfied, \textsc{NodeChoose} returns a null value. Indeed, the selection criteria given above effectively helps \textsc{Traverse} to efficiently obtain a domain of $Q_c$ with high quality. \looseness=-1 


\eat{
\begin{enumerate}
    \begin{enumerate}[(i)]
        \item Finding the neighbors, denoted as ($H_c=\{v_h\}$), each of which satisfies {\color{red}$v_h \in S_d$} and $v_h \sim u_{|S_d|}$ where $u_{|S_d|} \in D(Q_p)$;
        \item Choosing a node $v$ for further \textsc{Traverse} according to the following conditions (\textsc{NodeChoose}),
        \begin{equation}
            v := \left\{
            \begin{array}{lr}
                v_h \in H_c/ D(Q_c),  &  H_c/D(Q_c)\neq \varnothing\\
                v_h \in H_c\cap D(Q_c), & \ \ H_c/D(Q_c)= \varnothing, S_d \not\subseteq D(Q_c), c<m\\
                null, & {\rm otherwise}
            \end{array}.
            \right.
        \end{equation}
        Note that if the second condition holds, $c$ goes up by 1, indicating that the second statement will be performed no more than $m$ times;
        \item If $v$ is $null$, \te returns back to the previous level of the recursion; otherwise, \textsc{Traverse} pushes $v$ into the stack $S_d$, and executes a branch according to the condition $|S_d|=|V_p|$: {\color{red} calling \textsc{Expand} to try updating $D(Q_c)$ with $S_d$ if it is true}, or going to the next level of the recursion if it is false.
    \end{enumerate}
\end{enumerate}
}

\eat{  
\noindent (III) Starting from a valid node $v$, \textsc{Traverse} proceeds to traverse by referencing $S_d$ (lines 13-21). During the traversal, it first updates $c$ as $c+1$ if $v$ is picked from $H_c\cap D(Q_c)$ (line 14). It keeps detecting a discriminant condition $|S_d|$==$|V_p|$ (line 16), and invokes procedure \textsc{Expand} to {\color{red} try to update $D(Q_c)$} (line 17) if the condition is satisfied. Otherwise, \textsc{Traverse} invokes itself for deeper exploration (line 18). Afterwards, \textsc{Traverse} pops the upper-most node of $S_d$ (line 20), and picks a different node with \textsc{NodeChoose} for next round exploration. 
}

\begin{example}
\label{alg-running-2}
Continuing Example~\ref{alg-running-1}, \te restores stack ${S_d}$ by pushing $v_4$ onto it and calls \textsc{Traverse} to update $D(Q_2)$ in a depth-first manner. Firstly, \textsc{ConsExtr} is invoked. It identifies an {\em unseen} edge $e_u = (u_0, u_1)$, and obtains a set $H_c = \{v_0, v_1, v_2\}$, as $v_4 \in D_0(Q_1)$ and $v_i\sim u_1$ ($i\in [0,2]$). Afterwards, \traverse calls \textsc{NodeChoose} to pick a node for further exploration, and $v_0$ is chosen due to Condition A. \traverse next calls itself for traverse at a deeper level.  \looseness=-1
\eop
\end{example}

\vspace{0.5ex}
\noindent $\mathbf{Stage (III)}$: Starting from a valid node $v$, \textsc{Traverse} proceeds by referencing $S_d$ (lines 13-21). During the traversal, it first updates $c$ as $c+1$ if $v$ is picked from $H_c\cap D(Q_c)$ (line 14). Then, it keeps detecting a discriminant condition $|S_d|$==$|V_{p_p}|$ and invokes \textsc{Expand} to update $D(Q_c)$ if the condition is satisfied (lines 16-17). \looseness=-1

Procedure \textsc{Expand} (not shown) works as follows. If $Q_c$ is generated with $e_x=(u_i,u_j)$ via {\em forward expansion}, \textsc{Expand} searches neighbors $v'$ of $v$, where $v\in S_d$, $v\sim u_i$ and $v'\sim u_j$, and puts $v'$ along with nodes in $S_d$ in corresponding domain of $D(Q_c)$, as these nodes together already form valid matches of $Q_c$. For {\em backward expansion}, 
\textsc{Expand} verifies whether nodes in $S_d$ already form a match of $Q_c$ by referencing $e_x$ and $D_i(Q_c)$, $D_j(Q_c)$. If true, it updates $D(Q_c)$ along the same line as that for {\em forward expansion}.

\begin{figure}[t]
\centering
\includegraphics[width=\linewidth]{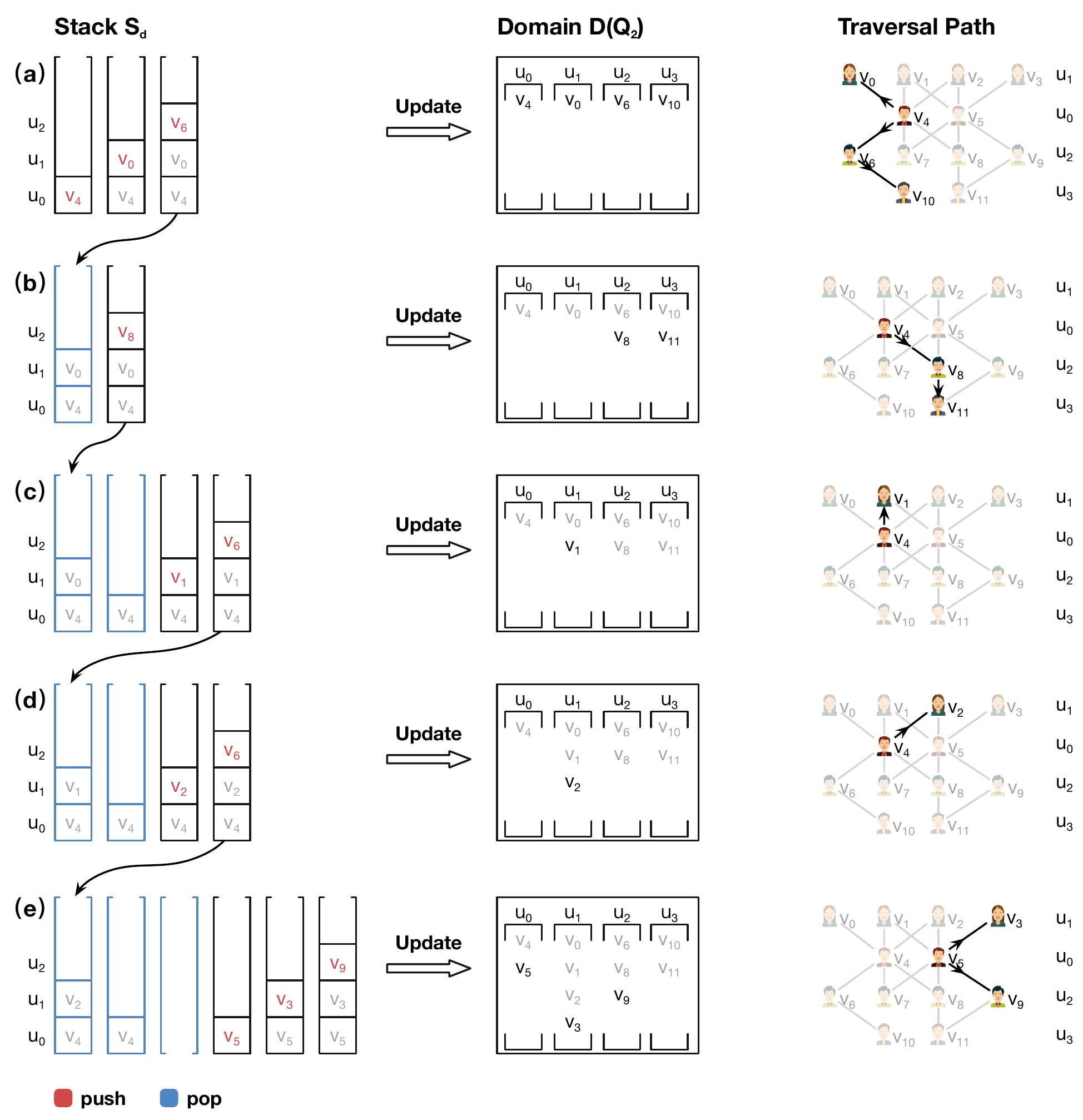}
\caption{\te Running Process}
\label{fig:RunProcess}
\vspace{-2ex}
\end{figure}

\begin{example}
\label{alg-running-3}
Recall Examples~\ref{alg-running-1} and \ref{alg-running-2}. At a deeper level, an {\em unseen} edge $e_u=(u_0, u_2)$ is used to guide traversal. Then $H_c = \{v_6, v_7, v_8\}$ is obtained as they are neighbors of $v_4$ that is in $S_d$. \textsc{NodeChoose} then picks $v_6$ (Condition A) and pushes it onto $S_d$. The status of $S_d$ is shown in Fig.~\ref{fig:RunProcess} (a) (right most stack). 
At this moment, \textsc{Expand} is invoked, as the discriminant condition $|S_d| == |V_{p_p}|$ is satisfied. \textsc{Expand} identifies a node $v_{10}$ (as the neighbor of $v_6$) such that $v_{10} \sim u_3$ of $Q_2$, according to the forward edge $(u_2,u_3)$. Then, a valid match of $Q_2$ forms. \textsc{Expand} puts $v_{10}$ as well as nodes in $S_d$ in corresponding columns of $D(Q_2)$. The upated $D(Q_2)$ is depicted in Fig.\ref{fig:RunProcess}(a). \looseness=-1
\eop
\end{example}

\eat{
\ie $Q_c$ is expanded with a new edge $(u,u')$, where $u'$ is a newly included node, \textsc{Expand} search neighbors $H_c =\{v'\}$ of $v$, where $v \sim u$ is in $S_d$ and $v'\sim u'$. If $H_c \neq null$, put those nodes (include nodes in $S_d$) into corresponding column of $D(Q_c)$. 
For backward expansion, \textsc{Expand} verifies backward edge Constraints to checks whether nodes in $S_d$ can form a match of $Q_c$. If true, updates $D(Q_c)$ with such a match by putting nodes of the match into corresponding column of $D(Q_c)$ (line 7).
}

Otherwise, \textsc{Traverse} invokes itself for deeper exploration (line 18). Afterwards, \textsc{Traverse} pops the upper-most node of $S_d$ (line 20), and picks a different node with \textsc{NodeChoose} for next round exploration (line 21).

\begin{example}
\label{alg-running-4}
Following previous examples, \textsc{Traverse} pops $v_6$, which is the upper-most node, out of $S_d$ and picks $v_8$ (Condition A) for next round iteration. Figure~\ref{fig:RunProcess} (b) shows how $S_d$ is changed, where $S_d$ after popup operation is colored in blue. As $|S_d|$ equals to $|V_{p_p}|$ now, \textsc{Expand} found that a new match can be formed with $v_{11}$ and updates $D(Q_2)$ with the new match. The updated $D(Q_2)$ is also shown in the middle of  Fig.~\ref{fig:RunProcess} (b).

One may refer to Fig.~\ref{fig:RunProcess} for the entire process of \te, where detailed changes of stack $S_d$, domain $D(Q_2)$, and traversal paths are provided. Detailed explanation is omitted due to space constraint. \eop
\end{example}

\eat{
\etitle{Correctness}. The correctness of \aprtopk is warranted by the following observations. (1) The pattern generation scheme will never miss any qualified pattern. (2) The estimation technique employed by \aprtopk guarantees to return a lower bound of support for a pattern. (3) The termination condition (Prop.~\ref{prop-early-termination}) ensures that when \aprtopk terminates, there exists no better patterns. \looseness=-1

\etitle{Complexity}. For a pattern $Q_c$ with node set $V_{p_c}$, there exist at most $|V|^{|V_{p_c}|}$ matches in a graph $G$. Thus, it takes $O(|V|^{|V_{p_c}|+1})$ time to identify matches of a candidate pattern $Q_{c}'$ ($Q_c\sqsubseteq Q_c'$). If $Q_c$ is a ``tree'' pattern, at most $|V_{p_c}||{\cal L}|$ candidate patterns can be generated after {\em forward expansion}; otherwise, at most $O(|V_{p_c}|^2)$ candidate patterns will be produced through {\em backward expansion}. Meanwhile, it still 
takes $O(|V|^{|V_{p_c}|+1}\cdot(|V_{p_c}|+1)^{|V_{p_c}|+1})$ time to verify whether a candidate $Q_c$ has been generated before, as at most $|V|^{|V_{p_c}|+1}$ candidate patterns may be generated earlier and each round isomorphism checking needs $O((|V_{p_c}|+1)^{|V_{p_c}|+1})$ time. 
By induction, it takes \aprtopk~$\sum_{i\in [1,n]} (|V_{p_c}|+i-1)|V|\cdot |V|^{|V_{p_c}|+i}(1+(|V_{p_c}|+i)^{|V_{p_c}|+i})$ times to verify all the subsequent patterns for a candidate $Q_c$, where $n$ refers to the expansion times and is bounded by $|V|$, $|{\cal L}|$ and $|V_{p_c}|$ are bounded by $|V|$. Since the iteration starts from single-edge patterns $Q_c$ and at most $|V|^2$ different $Q_c$ exists, 
thus \aprtopk is bounded by $O(|V|^{n}\cdot n^{n+1})$ time. \looseness=-1


The analysis above completes the proof of Theorem~\ref{theo-alg}. \eop
}

\eat{
Thus, starting with a stack containing only one node $v\in D_0(Q_p)$, \textsc{Traverse} generates matches in a depth-first manner and adds nodes from valid matches into $D(Q_c)$. Furthermore, based on the \textsc{Traverse} procedure, the iteration (lines 2-7) in \te will be repeated $|D_0(Q_p)|$ times, each of which starts with a node in $D_0(Q_p)$ and detects a match of $Q_c$ to update $D(Q_c)$. When the condition $|D_0(Q_c)|+|D_0(Q_p)|<\theta$ holds, the iteration breaks, indicating that $Q_c$ is an infrequent pattern (lines 6-7). After the iteration is finished, the support of $Q_c$ with $D(Q_c)$ will be calculated and returned as the final result (lines 8-9). 
}

\eat{
\begin{figure}[tb!]
\begin{center}
{\small
\begin{minipage}{4.5in}
\myhrule
\vspace{-1ex}
\vspace{-4.5ex}
\mat{0ex}{

\eat{
{\bf Procedure} \te~\\
{\sl Input:\/} a pattern $Q_c$, a candidate Domain $D_s$, a frequent threshold $\theta$, and a user\\ \hspace{6.5ex} defined parameter $M$ ($0 < M < 1$). \\
{\sl Output:\/} The \mni of pattern $Q_c$. \\
\bcc \hspace{1ex} initialize a stack $S_d:=\varnothing$ for storing match;  \\
\icc \hspace{1ex} \For each \kw{AS} in Domain \Do; \\
\icc \hspace{3ex}   start \kw{DFS} from head fraction and constantly execute \kw{DFS}: \\
\icc \hspace{5ex}       $Set_0 :=$ find all constraint nodes according to \kw{InfoList}; \\
\icc \hspace{5ex}       Divide $Set_0$ into $Set_1$ and $Set_2$ according to whether the node is already\\
\hspace{9ex} in $Q_c$.Domain; \\
\icc \hspace{5ex}       $Set := Set_2 \cup$ random pick no more than m nodes in $Set_1$; \\
\icc \hspace{5ex}       Set nodes in $Set$ as candidate traversal nodes; \\
\icc \hspace{5ex}       $S_d$ push the node that \kw{DFS} visited in current Iteration; \\
\icc \hspace{5ex}           \If the nodes in stack $S_d$ can form a match of pattern $Q_c$ \Then \\
\icc \hspace{7ex}          put the nodes into the corresponding columns of $Q_c$.Domain; \\
\icc \hspace{3ex}  $size_1:=$ size of $v_0$ Domain of $Q_c$; \\
\icc \hspace{3ex}  $size_2:=$ size of untraveled AS; \\
\icc \hspace{3ex}  \If  $size_1 + size_2 < \theta * M$  \Then \Break; \\
\icc \hspace{0ex} \Return \mni($Q_c$); \\
}
}

{\bf Procedure}~\textsc{Traverse}opt\\
{\sl Input:\/} A pattern $Q_c$, a candidate Domain $D_s$, a start node $n$.\\
{\sl Output:\/} Updated pattern $Q_c$. \\
\bcc \hspace{1ex} initialize a stack $S_d:=\varnothing$ for storing matches, a stack $S_t:=\varnothing$ for storing candidate traverse nodes;  \\
\icc \hspace{1ex} $S_t$.push($n$); \\
\icc \hspace{1ex} \While($S_t$ != NULL): \\
\icc \hspace{3ex}   $tn$ := $S_t$.pop(); \\
\icc \hspace{3ex}   $S_d$.push($tn$); \\
\icc \hspace{3ex}   \If $S_d$.size() == $D_s$.size() \Then \\
\icc \hspace{5ex}       \If the nodes in stack $S_d$ can form a match of pattern $Q_c$ \Then \\
\icc \hspace{7ex}          put the nodes into the corresponding columns of $Q_c$.Domain; \\
\icc \hspace{5ex}      $S_d$.pop(); \\ 
\icc \hspace{2ex}   \Else \\
\eat{
\icc \hspace{4ex}   $info$ := \kw{InfoList}.get($S_d$.size()); \\
\icc \hspace{4ex}   $nodes$ := get constraint nodes according to $info$ (node consistency and arc consistency);\\}
\icc \hspace{4ex}   $nodes$ := getConstraintNeighbors($Q_c$, $S_d$.size());\\
\icc \hspace{4ex}   $nodes$ := $nodes$ / $S_d$; \\
\icc \hspace{4ex}   Divide $nodes$ into $Set_1$ and $Set_2$ according to whether the node is already \\
in $Q_c$.Domain;\\
\icc \hspace{4ex}   $Set :=$ $Set_2 \cup$ random pick no more than $m$ nodes in $Set_1$; \\
\icc \hspace{4ex}   \If $Set$ ==NULL \Then \\
\icc \hspace{6ex}   $S_d$.pop(); \\
\icc \hspace{4ex}   \Else \\
\icc \hspace{6ex}       $S_t$.push($Set$); \\
\icc \hspace{0ex} \Return updated pattern $Q_c$; \\

\vspace{-5ex}
\myhrule
\end{minipage}
}
\end{center}
\vspace{-2ex}
\caption{Algorithm \te} \label{alg:TE}
\vspace{-3ex}
\end{figure}
}

\eat{
\begin{figure}[tb!]
\begin{center}
{\small
\begin{minipage}{4.5in}
\myhrule
\vspace{-1ex}
\vspace{-4.5ex}
\mat{0ex}{
{\bf Procedure}~\recursion~\\
{\sl Input:\/} a pattern $Q_c$, a candidate Domain $D_s$, an Onehop map \kw{OM}.\\
{\sl Output:\/} The \mni of pattern $Q_c$. \\
\bcc \hspace{1ex} initialize a stack $S_{t}:=\varnothing$ for DFS traveling; a stack $S_d:=\varnothing$ for storing match;  \\
\icc \hspace{1ex} \For each node $n$ in $D_s$[0] \Do \\
\icc \hspace{3ex}   $S_{t}$.Push($n$); \\
\icc \hspace{3ex}   \While (! isEmpty($S_t$)) \\
\icc \hspace{5ex}   $v:=S_t$.Pop();\\
\icc \hspace{5ex}   \If $S_d$.size == $D_s$.size \Then \\
\icc \hspace{7ex}       \If nodes in $S_d$ can form a match of $Q_c$ \Then  \\
\icc \hspace{9ex}          put the nodes into the corresponding columns of $Q_c$.Domain; \\
\icc \hspace{7ex}   pop the node from $S_d$ until it can push the node $v$; \\
\icc \hspace{4ex}   $S_d$.Push($v$); \\
\icc \hspace{4ex}   $Set_a:=$\kw{getFraction}($S_d$, $Q_c$.\kw{InfoList}, \kw{OM}) / $S_d$;\\
\icc \hspace{4ex}   $Set_u:=$ $Set_a$ / $Q_c$.Domain; \\
\icc \hspace{4ex}   $Set:=Set_u$ $\bigcup$ randomly pick no more than m nodes in ($Set_a$ / $Set_u$); \\
\icc \hspace{4ex}   $S_t$.Push($Set$); \\
\icc \hspace{0ex} \Return \mni($Q_c$); \\
}

\vspace{-5ex}
\myhrule
\end{minipage}
}
\end{center}
\vspace{-2ex}
\caption{Algorithm \te} \label{alg:TE}
\vspace{-3ex}
\end{figure}

\stitle{Algorithm}.
\te takes a pattern $Q_c$, a candidate Domain $D_s$ and an Onehop map \kw{OM} as input and outputs the \mni of pattern $Q_c$.
The algorithm traverses all ASs referring the Forward info in \kw{InfoList} in a depth-first manner to dynamically update the $Q_c$.Domain. 

\te first initializes an empty stack $S_t$ for DFS traveling and an empty stack $S_d$ for storing the embedding generated by the traversal process (line 1). 
It is worth noting that $S_d$[0] can only store the nodes in $f_0$, and $S_d$[1] can only store the nodes in $f_1$ and similar to $S_d$[n].
The algorithm traverses all ASs referring the Forward info in \kw{InfoList} in a depth-first manner to dynamically update the $Q_c$.Domain (lines 2-16). (where, lines 3, 4, 5, 14 are related to DFS operations.)
The algorithm regards $v$ ($v:= S_t$.Pop()) as the current traversed node (line 5) and puts node $v$ into stack $S_d$ to generate embeddings (line 10).
After that, the algorithm will get needed next fraction nodes set $Set$ in AS and push it into $S_t$ (lines 11-14). 
In the process, when the size of stack $S_d$ reaches $D_s$.size, algorithm will judge whether a match of pattern $Q_c$ can be formed by the nodes in stack $S_d$. If true, algorithm will put the nodes of match into corresponding column in $Q_c$.Domain (lines 6-8). Then stack $S_d$ will pop nodes preparing for pushing new visited node of \kw{DFS}.
}

\eat{
\begin{example}
\label{exa-recursion1}

As shown in Fig.\ref{fig:RunProcess}(a), \te restores the stack ${S_d}$ by pushing $v_4$.
Afterwards, \te calls \textsc{Traverse} to search into the next level.
In this level, \textsc{ConsExtr} identifies the current expand edge $e_u = (u_0, u_1)$, then get the neighbors $H_c = \{v_0, v_1, v_2\}$ of $v_4$, where $v_4 \sim u_0$ and $\{v_0, v_1, v_2\} \sim u_1$.
Next, \traverse calls \textsc{NodeChoose} to pick one node for further step, so, $v_0$ (Condition A) is pushed into $S_d$ and \traverse calls itself into next level.
Here, expand edge $e_u = (u_0, u_2)$ and $H_c = \{v_6, v_7, v_8\}$ of $v_4$.
\traverse pick $v_6$ (Condition A) into $S_d$. In this time, \traverse calls \textsc{Expand}, due to the discriminant condition $|S_d| == |V_{p_p}|$ is satisfied.
As $Q_c$ is generated via forward expansion with a new edge $e_x = (u_2, u_3)$, \textsc{Expand} search the neighbors ($H_c := \{v_{10}\}$ in $G$) of $v_6$ ($v_6 \sim u_2$). As $H_c \neq null$, \textsc{Expand} put those nodes (include nodes in $S_d$) into corresponding column of $D(Q_2)$ (Fig.\ref{fig:RunProcess}(a)). 
Afterwards, $S_d$ pops a node and \traverse calls \textsc{NodeChoose}. $v_8$ is chosen (Condition A) and pushed into $S_d$ for further detection. $\{v_4, v_0, v_8, v_{11}\}$ is added into $D(Q_2)$ (Fig.\ref{fig:RunProcess}(b)).
Then, \textsc{NodeChoose} returns $null$ according to Condition C. Meanwhile, the while loop is terminated and back to last level.
Immediately, $S_d$ pops a node and \traverse calls \textsc{NodeChoose}, $v_1$ is chosen (Condition A) and pushed into $S_d$ for next level.
In this current, expand edge $e_u = (u_0, u_2)$ and neighbors $H_c = \{v_6, v_7, v_8\}$ of $v_4$ ($\{v_6, v_7, v_8\} \sim u_2$, $v_4 \sim u_0$).
\traverse calls \textsc{NodeChoose}, and $v_6$ is chosen to push into $S_d$ (Condition B) and c plus by one. Discriminant condition $|S_d|==|V_{p_p}|$ satisfied, \traverse calls \textsc{Expand} and $\{v_4, v_1, v_6, v_8\}$ is added into $D(Q_2)$ (Fig.\ref{fig:RunProcess}(c)).
Immediately, $S_d$ pops a node and \traverse calls \textsc{NodeChoose}. $null$ is returned by \textsc{NodeChoose}, since Condition B is not satisfied as $S_d \subseteq D(Q_2)$.
$S_d$ pops again and only $v_4$ in $S_d$.
\traverse calls \textsc{NodeChoose}, $v_2$ is chosen to push into $S_d$. Same to $v_1$, $\{v_4, v_2, v_6, v_8\}$ is added into $D(Q_2)$ (Fig.\ref{fig:RunProcess}(d)). When back to $S_d = \{v_4\}$ third time, \textsc{NodeChoose} return $null$, since all the nodes that can form a match with $v_4$ are added into $D(Q_2)$.
\te executes next iteration, restoring the $S_d$ by pushing $v_5$ after cleaning. Then \te calls \traverse for further detection.
In this level, \textsc{ConsExtr} identifies the current expand edge $e_u = (u_0, u_1)$, then get the neighbors $H_c = \{v_1, v_2, v_3\}$ of $v_5$, where $v_5 \sim u_0$ and $\{v_1, v_2, v_3\} \sim u_1$. \traverse calls  \textsc{NodeChoose} to get $v_3$ (the only node satisfied Condition A). Then \traverse pushes it into $S_d$ and calls itself into next level.
In this current, expand edge $e_u = (u_0, u_2)$ and $H_c = \{v_7, v_8, v_9\}$ of $v_5$. \traverse gets $v_9$ by calling \textsc{NodeChoose} (the only node satisfied Condition A) and pushes $v_9$ into $S_d$. \traverse executes \textsc{Expand}, as discriminant condition $|S_d| == |V_{p_p}|$ is satisfied.
$\{V_5, v_3, v_9, v_11\}$ is added into $D(Q_2)$ (Fig.\ref{fig:RunProcess}(e)). In this time, we can easily observe that no strange nodes in this path for $D(Q_2)$ (each level \textsc{NodeChoose} satisfies condition C), so it is no need for \traverse to visit nodes already in $D(Q_2)$ as it is no help to $\supp(Q_2)$. \eop

\eat{
and recursively find $v_0$, $v_6$ and push them into stack $\kw{S}$ 
In each iteration, \textsc{Traverse} initializes two empty stacks $S_d$ for storing matches and stack $S_t$ for maintaining the DFS traverse, first.
Let's look at the change process of stack $S_d$ in Fig.~\ref{fig:RunProcess}. In 00, $S_d$ first put $v_4$ a node from $u_0$ into stack $S_d$, then \textsc{Traverse} search next \img $u_1$ according to the structure of pattern $Q_1$, then find node $v_0$ and push it into $S_d$. repeatedly, \textsc{Traverse} find node $v_6$ and push it into $S_d$. In this time, the $S_d$ reach the size of pattern $Q_1$. \textsc{Traverse} 
executes \textsc{Expand} and verified that nodes in $S_d$ can form a match of $Q_1$ and extend a match of $Q_2$.
}

\eat{
First, the algorithm traverses $AS_0$. Push $C_0$ into stack $S_d$ and then traverse AS in a DFS manner according to the \kw{infoList}.
when the size of the stack is reached 4 (the size of the candidate Domain $D_s$), the algorithm will judge whether the nodes in the stack can form a match of the child pattern $Q_c$, and add nodes that can form a match of the pattern $Q_c$ to Domain.
The stack changes are shown in Fig.~\ref{fig:RunProcess}(a), Among them, lines 01-18 show the complete traversal process of $AS_0$, and lines 19-44 show the complete traversal process of $AS_1$. 
It is worth mentioning that the original recursion can output all matches corresponding to the pattern.

In Fig.~\ref{fig-Auxiliary}(c), we can easily find that there are overlaps between different \kw{AS}s, which makes it possible to find new matches of the pattern in the traversal process of origin \te, but it does not make Domain get new nodes.
In our approximation method, the algorithm's access to a part of the repeated path is restricted. (shown in Fig.~\ref{alg:TE} lines 5-6)
Algorithm will get process like  Fig.~\ref{fig:RunProcess}(c). The difference between origin \te and \te is The latter will not traverse all the node that have been found. It just random pick m(here suppose m=1) nodes already in $Q_c$.Domain as DFS candidate to reduce the time complexity of calculation. As Fig.~\ref{fig:RunProcess}(d) showing, algorithm only visited one of the repeated paths.
}

\end{example}
}

\stitle{Remarks}. (1) The parameter $m$ controls total number of nodes used for exploration. As verified via experimental studies, guided traversal with limited numbers substantially improves efficiency of mining process, taking only 11.7\% time and 31.5\% memory of its counterpart, while obtaining 100\% recall. 
(2) When performing support estimation for a candidate pattern $Q_c$, \te leverages the cached domain of $Q_p$ (as the ``parent'' of $Q_c$ and computed in earlier), which substantially improves efficiency. 
(3) Note that the decision problem of \textsc{TopkPM} is already NP-hard, so no matter how desired, the \textsc{TopkPM} problem can not be solved in PTIME. Despite high computational cost, \aprtopk works more efficiently than its counterparts, owing to its approximation scheme for supports estimation and early termination property. \looseness=-1

\eat{
\begin{example}
\label{exa-alg-tree}
On graph $G$ of Fig.~\ref{fig-running-example} (a), We suppose support threshold $\theta = 3$. \aprtopk first finds out the frequent edges and converts them into edge patterns $Q_1$-$Q_5$, as their supports all equal to $3$. Then, \aprtopk applies \kw{ForwardTree}(resp. \kw{BackwardTree}) to generate candidate patterns, in a {\em level-by-level} manner. For example, using pattern $Q_1$, \aprtopk generates candidate patterns by enlarging $Q_1$ with other frequent single-edge patterns and produces $L =\{Q_{11},Q_{12},Q_{13},Q_{14},Q_{15}\}$. Four levels candidate patterns are shown in Fig.~\ref{fig-running-example} (c), where the red arrow represents the new forward edge generated during the process of \kw{ForwardTree}, and the blue arrow represents the new backward edge generated during the process of \kw{BackwardTree}.
\end{example}
}

\eat{
\begin{figure}[tb!]
\begin{center}
{\small
\begin{minipage}{4.5in}
\myhrule
\vspace{-1ex}
\mat{0ex}{
{\bf Procedure}~\uplevel~\\
{\sl Input:\/} \= A candidate patterns set $L$, a tree ${\cal T}$.  \\
{\sl Output:\/} $L$ that has been pruned some infrequent patterns. \\
\bcc \hspace{1ex} initialize $\lu:=\varnothing$; $\kw{L_{t}}:=\varnothing$; \\
\icc \hspace{1ex}   \For \Each pattern $Q_c$ in $L$ \Do \\
\icc \hspace{3ex}   \For \Each vertex $v$ in $Q_c$ \Do \\
\icc \hspace{5ex}   \If vertex $v$ is not the extending vertex and the subpattern $Q_s$ after $Q_c$ \\
\hspace{9.5ex} removes $v$ is connecting \Then \\
\icc \hspace{7ex}   $\kw{L_{t}}:=\kw{L_{t}} \bigcup \{Q_s\}$; \\
\icc \hspace{3ex}  \If ${\cal T}$ does not contain all patterns in \kw{L_{t}} \Then \\
\icc \hspace{5ex}   $\lu:=\lu \bigcup \{Q_c\} $;\\
\icc \hspace{3ex}   reset $\kw{L_{t}}:=\varnothing$; \\
\icc \hspace{1ex} $L := L/\lu$; \\
\icc \hspace{1ex} \Return $L$; \\
}

\vspace{-5ex}
\myhrule
\end{minipage}
}
\end{center}
\vspace{-2ex}
\caption{The Pseudo-code of \uplevel} \label{alg:upleve}
\vspace{-3ex}
\end{figure}
}


\eat{
\subsection{{\color{red}Optimizations}}
 To further improve the performance of \aprtopk, we introduce some optimization techniques used in programming.
 
\subsubsection{Upper level pruning} Upper level pruning can prune infrequent candidates without support verification. Recall the {\em anti-monotonic} property of our support metric. It ensures that if a pattern $Q$ is not frequent, then any pattern that subsumes $Q$ is not frequent as well. Based on this property, given a candidate pattern $Q_c$, one only needs to verify the frequency of each sub-pattern $Q_c'$ of $Q_c$. This can be achieved by checking whether each $Q_c'$ is already in ${\cal T}$, since ${\cal T}$ is generated following a  ``level-wise'' strategy, hence before $Q_c$ is generated, all of its frequent sub-patterns must be already in ${\cal T}$. 

\subsubsection{One-Hop Map} 
\te will frequently query the neighbors of a certain node, so for the relationship between nodes and their neighbors, a cache can greatly improve the efficiency of program execution.
We store frequent edges in the form of map to speed up \traverse to obtain constrained neighbor nodes. When executing frequent pattern mining in single large graph, after frequent threshold $\theta$ is fixed, most of the edges are excluded. So, a few of edges need to be recorded into one-hop map and it won't cause too much burden on memory. Experiments also verified this.
}

\eat{
\begin{example}
\label{exa-uplevelpruning}
In order to identify whether a candidate pattern must be infrequent, at first \uplevel pick a candidate pattern $Q_{131}$ from the level ($h=2$) of Fig.~\ref{fig-running-example} (c). Then we only take out one vertex that is not marked with red each iteration, as a result we can split $Q_{131}$ into $Q_{11}$ and $Q_{13}$. Next we check whether the level ($h=1$) of ${\cal T}$ in Fig.~\ref{fig-running-example} (d) contains these two patterns. Since ${\cal T}$ only contains $Q_{13}$, it can be judged that $Q_{131}$ must not be frequent according to the anti-monotonic rule.
\end{example}
}

\eat{
\stitle{Algorithm}. \sverify takes a candidate pattern $Q_c$, a candidaie Domain $D_s$, a support threshold $\theta$ and a minimum ratio $M_r$ as input and outputs true, if $Q_c$ is a frequent pattern of $G$; false, oterwise. \looseness=-1

\sverify first get the approximate $supp$ of $Q_c$ through procedure \te (line 1). Then we judge whether the $supp$ of $Q_c$ is greater than the threshold $\theta$, if it is greater than the threshold, \sverify will return true (line 2). Since the $Q_c$.Domain obtained by \te is an approximate result, we need to consider the matches of $Q_c$ that may be missed. Now we give a maximum difference range $M_r$. If the $supp$ of $Q_c$ is greater than $\theta-M_r$ and less than $\theta$, then we think that a second verification of $Q_c$ is needed to make up for the error (line 3). Since the second verification is for those matches of $Q_c$ that may be missed, we need to find those sets in $Q_c$ whose size is less than $\theta$ and remove the same nodes from the candidate set \kw{D_s}. Through the above pruning, not only those matches that may be missed are excavated, but also reduced the times of traversals in the mining process (lines 4-6). Then, we execute procedure \te to update $Q_c$.Domain list (line 7). If the $supp$ of $Q_c$ is greater than the threshold $\theta$, \sverify will return true (line 8). If the size of $Q_c$.Domain[$i$] is still less than the threshold $\theta$ after the second verification, then we return false (line 9). \looseness=-1 

\etitle{early judgment}. When Travel embedding has been unable to find matches or rarely finds matches, the search space will be difficult to prune. At this time, the calculation in the associative space is equivalent to exhaustion.
Therefore, we have introduced early judgment, that is, in the calculation process of \te,
If satisfied $Nodes_{t} + Nodes_{u} < \theta * M_r$,
directly interrupt and judge that the pattern is not frequent (where $Nodes_{t}$ denotes the nodes can form a match, $Nodes_{u}$ denotes the nodes not traversed, frequent threshold $\theta$ and minimum rate $M_r$).

\begin{figure}[tb!]
\begin{center}
{\small
\begin{minipage}{4.5in}
\myhrule
\vspace{-1ex}

\mat{0ex}{
{\bf Procedure}~\sverify\\
{\sl Input:\/} \= a pattern $Q_c$, a candidate Domain $D_s$, support threshold $\theta$, \\
\hspace{6.5ex} a maximum difference range $M_r$. \\
{\sl Output:\/} \textbf{true}, \If $Q_c$ is a frequent pattern of $G$; \textbf{false}, otherwise. \\
\bcc \hspace{1ex} $supp:=$ $\te(Q_c,D_s,\theta)$; \\
\icc \hspace{1ex} \If $supp$ $\geqslant$ $\theta$ \Return \textbf{true}; \\
\icc \hspace{1ex} \Else \If $supp$ $<$ $\theta$ and $supp$ $\geqslant$ $(\theta-M_r)$ \Then \\
\icc \hspace{3ex} \For \Each column $i$ in $Q_c.$Domain \Do \\
\icc \hspace{5ex} \If $Q_c.$Domain[$i$].size $<$ $\theta$ \Then \\
\icc \hspace{7ex} $D_s'$ $:=$ $D_s$ $-$ $Q_c.$Domain$[i]$; \\
\icc \hspace{7ex} $supp:=$ $\te(Q_c,D_s',\theta)$; \\
\icc \hspace{7ex}   \If $supp$ $\geqslant$ $\theta$ \Then \Return \textbf{true}; \\
\icc \hspace{7ex}   \Else \If $Q_c.$Domain$[i]$.size $<$ $\theta$ \Then \Return \textbf{false};\\
\icc \hspace{1ex} \Else \Return \textbf{false}; \\
}

\vspace{-5ex}
\myhrule
\end{minipage}
}
\end{center}
\vspace{-2ex}
\caption{The Pseudo-code of \sverify} \label{alg:isfrequent}
\vspace{-3ex}
\end{figure}
}

\eat{
\stitle{Optimizations}. Simply sending all the $\~{G}$ to $S_c$ may incur excessive data shipment, since there may exist a large number of partial matches. 
To address the issue, we develop an optimization strategy to extract the substructure from a {\em partial match} $G_s$. We first compute the topological rank of $G_s$. This is doable since both $G_s$ and its corresponding pattern $Q_c'$ are ``trees''. Fixing $r(v)=0$ for each order node $V$ in $G_s$, we adjust the topological ranks of the remaining nodes by ... (To be continued.)
}


\section{Experimental Study}
\label{sec-expt}

Using real-life graphs and synthetic data, we conducted comprehensive experimental studies to evaluate: efficiency, memory cost, effectiveness (recall) and scalability of our algorithm \aprtopk, compared with baseline methods. 

\subsection{Experimental setting}

\stitle{Real-life graphs}. We used three real-life graphs: 
(a) {\em Amazon}~\cite{amazon-data}, a product co-purchasing network with $0.41$ million nodes and $3.35$ million edges. 
(b) {\em Mico}~\cite{GraMi2014}, a dataset models the Microsoft co-authorship information with $0.1$ million nodes and $1.08$ million edges. 
(c) {\em Youtube}~\cite{youtube}, a network of videos and their related videos from {\em Youtube} with $0.15$ million nodes and $1.05$ million edges.

\stitle{Synthetic graphs}. We also designed a generator to produce synthetic graphs $G=(V, E, L)$, controlled by the numbers of nodes $|V|$ and the number of edges $|E|$, where
$L$ is taken from an alphabet of $1K$ labels. We generated synthetic graphs following
the evolution model~\cite{Garg09}: an edge was attached
to the high degree nodes with higher probability. 
The size of $G$ is up to $0.5$ million nodes and $5$ million edges.



\stitle{Implementations}. We implemented algorithm \aprtopk and the following counterparts, all in Java. 

\begin{itemize}
    \item GRAMI, which identifies frequent patterns with the algorithm in \cite{GraMi2014}, ranks patterns based on our interestingness metric and selects top-$k$ ones. \looseness=-1
    \item  AGRAMI, the approximate version of GRAMI. Along the same line as GRAMI, AGRAMI first discovers frequent patterns with the approximate version of GRAMI. During this period, it sets the time-out to occur after $f(\alpha)$ iterations of the search, where $f(\alpha)=\alpha^{n}\prod_{1}^{n}\left|D_{i}\right|+\beta$, $\alpha\in (0,1]$ is a user-defined parameter, $\beta$ is a constant, $D_{i}$ is the image $\img(u_i)$ of pattern node $u_i$ 
    and $n$ is the number of pattern nodes. In this way, AGRAMI achieves better efficiency, at the cost of missing false negatives. After frequent patterns are discovered, the top-$k$ pattern selection is processed in the same way as GRAMI. \looseness=-1
\end{itemize}

In our test, the testbed includes a machine with 2.3 GHz CPU and 16 GB RAM, running JDK v11.0.9 on Windows 10. Each test was run five times and the average is reported.

\stitle{Parameters}. For \aprtopk, we fixed parameter $m=2$ (used in procedure \textsc{NodeChoose} ). For \agrami, we fixed $\alpha$ as $2*10^{-5}$, $2*10^{-4}$, $7*10^{-3}$ and $10^{-5}$ on {\em Amazon}, {\em Mico}, {\em Youtube} and synthetic graphs, respectively.


\eat{
\begin{figure*}[tb!]
\begin{center}
\subfigure[Response time ({\em Amazon})]{\label{fig-amazon-t-rt}
{\includegraphics[width=4.cm,height=2.35cm]{fig/Runtimeofall_amazon.eps}}} \hfill
\subfigure[Response time ({\em Mico})]{\label{fig-twitter-t-rt}
{\includegraphics[width=4.cm,height=2.35cm]{fig/Runtimeofall_Mico.eps}}} \hfill
\subfigure[Response time ({\em Youtube})]{\label{fig-facebook-t-rt}
{\includegraphics[width=4.cm,height=2.35cm]{fig/Runtimeofall_youtube.eps}}} \hfill
\subfigure[Memory cost({\em Amazon})]{\label{fig-amazon-m-all}
{\includegraphics[width=4.cm,height=2.35cm]{fig/Memoryofall_Amazon.eps}}} \hfill
\subfigure[Memory cost({\em Mico})]{\label{fig-twitter-m-all}
{\includegraphics[width=4.cm,height=2.35cm]{fig/Memoryofall_Mico.eps}}} \hfill
\subfigure[Memory cost({\em Youtube})]{\label{fig-facebook-m-all}
{\includegraphics[width=4.cm,height=2.35cm]{fig/Memoryofall_youtube.eps}}} \hfill
\subfigure[Recall ({\em Amazon})]{\label{fig-amazon-t-recall}
{\includegraphics[width=4.cm,height=2.35cm]{fig/Recallofall_amazon.eps}}} \hfill
\subfigure[Recall ({\em Mico})]{\label{fig-twitter-t-recall}
{\includegraphics[width=4.cm,height=2.35cm]{fig/Recallofall_mico.eps}}} \hfill
\subfigure[Recall ({\em Youtube})]{\label{fig-facebook-t-recall}
{\includegraphics[width=4.cm,height=2.35cm]{fig/Recallofall_youtube.eps}}} \hfill
\subfigure[Response time ({\em Amazon})]{\label{fig-amazon-k-rt}
{\includegraphics[width=4.cm,height=2.35cm]{fig/Runtimeoftopk_amazon.eps}}} \hfill
\subfigure[Response time ({\em Mico})]{\label{fig-twitter-k-rt}
{\includegraphics[width=4.cm,height=2.35cm]{fig/Runtimeoftopk_mico.eps}}} \hfill
\subfigure[Response time ({\em Youtube})]{\label{fig-facebook-k-rt}
{\includegraphics[width=4.cm,height=2.35cm]{fig/Runtimeoftopk_youtube.eps}}} \hfill
\subfigure[Memory cost({\em Amazon})]{\label{fig-amazon-m-top}
{\includegraphics[width=4.cm,height=2.35cm]{fig/Memoryoftop_amazon.eps}}} \hfill
\subfigure[Memory cost({\em Mico})]{\label{fig-twitter-m-top}
{\includegraphics[width=4.cm,height=2.35cm]{fig/Memoryoftop_mico.eps}}} \hfill
\subfigure[Memory cost({\em Youtube})]{\label{fig-facebook-m-top}
{\includegraphics[width=4.cm,height=2.35cm]{fig/Memoryoftop_youtube.eps}}} \hfill
\subfigure[Recall ({\em Amazon})]{\label{fig-amazon-k-recall}
{\includegraphics[width=4.cm,height=2.35cm]{fig/Recalloftop_amazon.eps}}} \hfill
\subfigure[Recall ({\em Mico})]{\label{fig-twitter-k-recall}
{\includegraphics[width=4.cm,height=2.35cm]{fig/Recalloftop_mico.eps}}} \hfill
\subfigure[Recall ({\em Youtube})]{\label{fig-facebook-k-recall}
{\includegraphics[width=4.cm,height=2.35cm]{fig/Recalloftop_youtube.eps}}} \hfill
\end{center}
\caption{Performance of \apm on real-life graphs} \label{fig-disrquery}
\end{figure*}
}

\begin{figure}
    \centering
    \includegraphics[width=\linewidth]{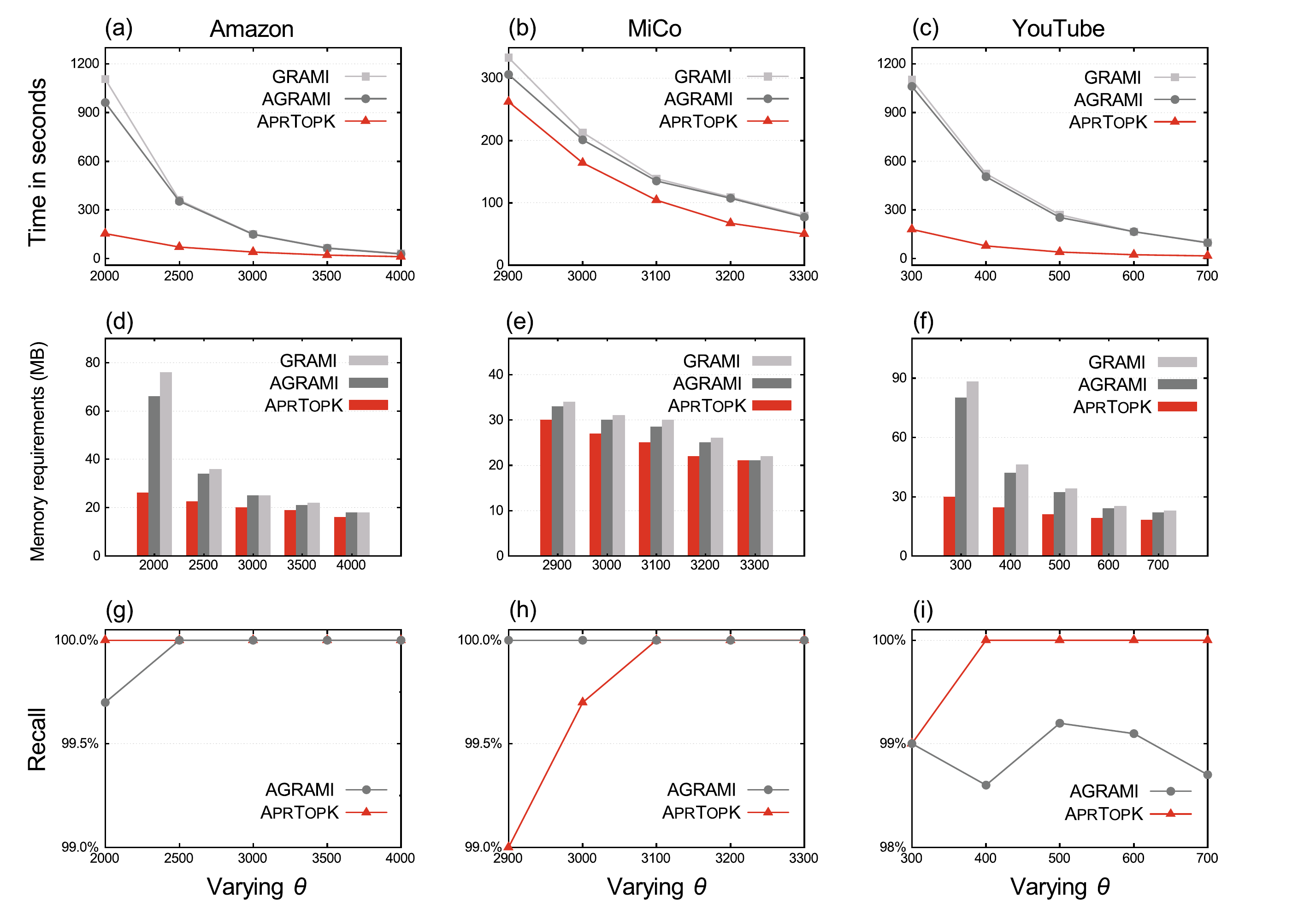}
    \caption{Exp-1: Influence of $\theta$}
    \label{fig:exp1}
\end{figure}

\begin{figure}
    \centering
    \includegraphics[width=\linewidth]{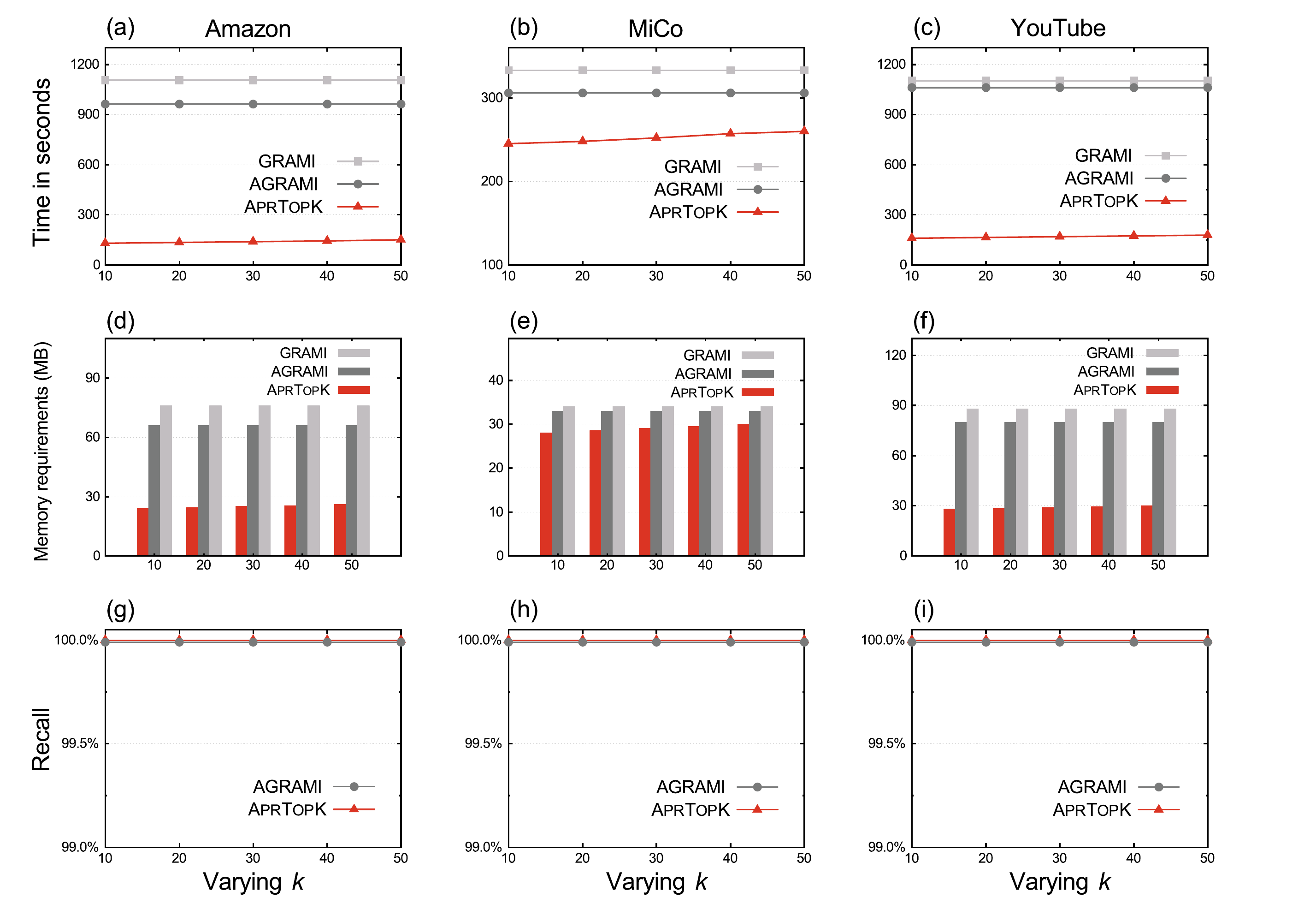}
    \caption{Exp-2: Influence of $k$}
    \label{fig:exp2}
\end{figure}

\subsection{Experimental results}

\stitle{Exp-1: Influence of $\theta$}. To see the influence of $\theta$, we fixed $k$ as a large number. It is a fair setting, since a complete (resp. incomplete but still large) set of frequent patterns need to be mined in \grami (resp. \agrami) in spite of the increase of $k$, while the increase of $k$ weakens \apm.


We then varied the support threshold $\theta$ from $2K$ to $4K$ in $0.5K$ increments, $2.9K$ to $3.3K$ in $0.1K$ increments and $0.3K$ to $0.7K$ in $0.1K$ increments on {\em Amazon}, {\em Mico} and {\em Youtube}, respectively. \looseness=-1 

\etitle{Efficiency}. Figures~\ref{fig:exp1}(a)-\ref{fig:exp1}(c)
report the response time of all the algorithms on {\em Amazon}, {\em Mico} and {\em Youtube}, respectively, which tell us the following. (1) With the increase of support threshold $\theta$, all the algorithms take shorter time, because fewer candidate patterns and their matches have to be verified. (2) \apm outperforms~\grami and \agrami in all cases and is less sensitive to the increase of $\theta$, since \apm is able to dramatically reduce the cost for candidate patterns verification. On {\em Amazon}, {\em Mico} and {\em Youtube} \apm only takes on average 17.4\%, 74.1\% and 15.8\% time of \grami, respectively. In particular, our algorithm takes only 13.8\% time of \grami, when $\theta = 2K$, while obtaining 100\% recall, on {\em Amazon}.

\eat{
of nodes existing in the \kw{Domain} of a candidate pattern, while \grami and \agrami need to perform a large number of repeated matching on the nodes that have been existed in the \kw{Domains}, thereby greatly increasing the number of iterations.}

\etitle{Memory cost}. Figures~\ref{fig:exp1}(d)-\ref{fig:exp1}(f)
show the memory footprint of the algorithms over {\em Amazon}, {\em Mico} and {\em Youtube}, respectively. We find that (1) the memory cost of all the algorithms drops with the increase of $\theta$, as fewer candidates need to be verified. (2) \apm consumes less memory than \grami and \agrami on three graphs, as expected; it incurs 59.6\%, 87.4\% and 52.3\% memory cost of \grami, on average, at {\em Amazon}, {\em Mico} and {\em Youtube}, respectively. 

\eat{
\etitle{Recall}. Figures~\ref{fig:exp1}(g)-\ref{fig:exp1}(i) show the recall of \apm and \agrami, \ie the ratio of patterns returned by \apm and \agrami vs. the complete set of frequent patterns. We find the following: (1) the recall of \apm is always higher than that of \agrami on three graphs. In particular, the recall of \apm even reaches 100\% on {\em Mico} (with $\theta>4K$ ) and {\em Youtube}. (2) Overall, when $\theta$ grows, the recall of both algorithms grows as well (not monotonically increasing). This is because, for a large $\theta$, the set of frequent patterns becomes smaller, which favors top-$k$ selection. (3) 
The recall of \agrami is influenced not only by $\theta$, but also by a set of parameters (\eg $\alpha$ etc.). We have tested \agrami with smaller $\alpha$ and find that its recall and efficiency mutually restrict. Due to space constraint, we omit details here. \looseness=-1}

\etitle{Recall}. Figures~\ref{fig:exp1}(g)-\ref{fig:exp1}(i) show the recall of \apm and \agrami, i.e., the ratio of patterns returned by \apm and \agrami vs. the complete set of frequent patterns. We find the following: (1) the recall of \apm is always higher than 99\% on three graphs. In particular, the recall of \apm even maintains 100\% on {\em Amazon}. (2) Overall, when $\theta$ grows, the recall of both algorithms grows as well (not monotonically increasing). This is because, for a large $\theta$, the set of frequent patterns becomes smaller, which favors top-$k$ selection. (3) 
The recall of \agrami is influenced not only by $\theta$, but also by a set of parameters (\eg $\alpha$ etc.). We have tested \agrami with smaller $\alpha$ and find that its recall and efficiency are mutually restricted. Due to space constraint, we omit details here. \looseness=-1

\vspace{0.5ex}

It is noted that, \te gets less efficiency and memory cost advantages on {\em Mico}.
The main reason lies in that the effectiveness of traversal strategy on {\em Mico} is not as good as that on {\em Amazon} and {\em Youtube}. Indeed, \te prefers to traverse from nodes that have not been seen since these unseen nodes can contribute the support of a pattern. This underlying feature results in that \te can achieve better performance to evaluate supports of those patterns whose corresponding matches have a large part of overlap (see Example~\ref{exa-alg-Domains}). As the graph structure of {\em Mico} does not favor \te very well from the perspective of match overlap, the performance gains from \te hence become less.

\stitle{Exp-2: Influence of $k$}. Fixing $\theta$ = $2K$, $2.9K$ and $0.3K$ for {\em Amazon}, {\em Mico} and {\em Youtube}, respectively, we varied $k$ from $10$ to $50$ in $10$ increments and compared \apm with \grami and \agrami. 

\etitle{Efficiency}. Results shown in Figures~\ref{fig:exp2}(a)-\ref{fig:exp2}(c) tell us the following. (1) \apm performs much more efficiently than \grami, owing to its approximation scheme employed by support estimation and {\em early termination} property. For example, \apm only takes on average 13.8\%, 78.6\% and 16.4\% time of \grami at {\em Amazon}, {\em Mico} and {\em Youtube}, respectively. (2) \apm is sensitive to the increase of $k$ since it has to verify more candidate patterns before the termination condition can be satisfied, while \grami and \agrami are not influenced by $k$ \wrt efficiency, as both of them apply the ``find-all-select'' strategy. 

\etitle{Memory cost}. Figures~\ref{fig:exp2}(d)-\ref{fig:exp2}(f) show the memory footprint of all the algorithms. We find the following. All the algorithms are not sensitive to the varying of $k$. The reasons are twofold: (1) \grami and \agrami are almost not influenced by the change of $k$, hence the memory requirements remain unchanged for both of them; and (2) \apm stops only when termination condition is satisfied, however a larger $k$ does not dramatically increase the memory cost, meanwhile we find that \apm consumes, on average, 34.2\% (resp. 88.2\%, 34.1\%) memory of \grami, on {\em Amazon} (resp. {\em Mico} and {\em Youtube}). 

\eat{
\etitle{Recall}. Figures~\ref{fig-amazon-k-recall}-\ref{fig-facebook-k-recall} report the recall of \apm and \agrami. We find the following. (1) \apm always performs better than \agrami. For example, the recall of \apm remains 100\% on both {\em Amazon} and {\em Youtube}, and exceeds 90\% on {\em Mico}. (2) The recall of \apm is insensitive to the change of $k$, because its  verification mechanism ensures that frequent patterns are hard to miss. }

\etitle{Recall}. Figures~\ref{fig:exp2}(g)-\ref{fig:exp2}(i) report the recall of \apm and \agrami. We find the following. Both \apm and \agrami perform very well on all datasets and the recall of them remains 100\% when $k$ increases from 10 to 50. 

\eat{
\begin{figure*}[tb!]
\begin{center}
\subfigure[Response time({\em Synthetic})]{\label{fig-rt-all}
{\includegraphics[width=4.cm,height=2.35cm]{fig/RuntimeofG.eps}}}  \hfill
\subfigure[Memory cost({\em Synthetic})]{\label{fig-rt-all}
{\includegraphics[width=4.cm,height=2.35cm]{fig/MemoryofG.eps}}}  \hfill
\subfigure[Recall ({\em Synthetic})]{\label{fig-recall-all}
{\includegraphics[width=4.cm,height=2.35cm]{fig/RecallofG.eps}}} \hfill
\end{center}
\vspace{-4ex}
\caption{Scalability of \apm} 
\label{fig-disrquery}
\end{figure*}
}
\begin{figure}[t]
    \centering
    \includegraphics[width=\linewidth]{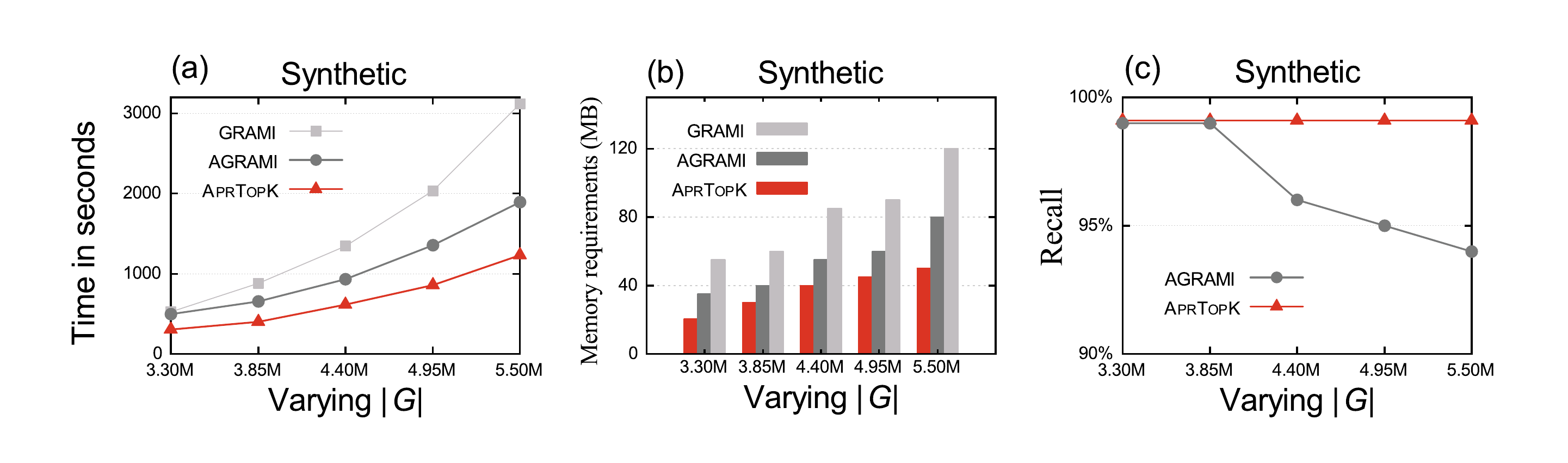}
    \vspace{-2ex}
    \caption{Exp-3: Scalability}
    \label{fig:my_label}
    \vspace{-2ex}
\end{figure}

\stitle{Exp-3: Scalability}. Fixing $\theta$ = $1K$ and $k$ = 50, we varied $|G|$ from $(0.3M,3M)$ to $(0.5M,5M)$ with $0.05M$ and $0.5M$ increments on $|V|$ and $|E|$, respectively, and compared \apm with \grami and \agrami. As shown in Figures~\ref{fig:my_label}(a)-\ref{fig:my_label}(c), (1) all the algorithms take longer time and consume more memory on larger graphs, as expected; (2) \apm is less sensitive to $|G|$ than others, \wrt response time and memory footprint, showing its better scalability; and (3) \apm shows a more steady recall than \agrami, with the increase of $|G|$. \looseness=-1

\begin{figure}[t]
    \centering
    \includegraphics[width=\linewidth]{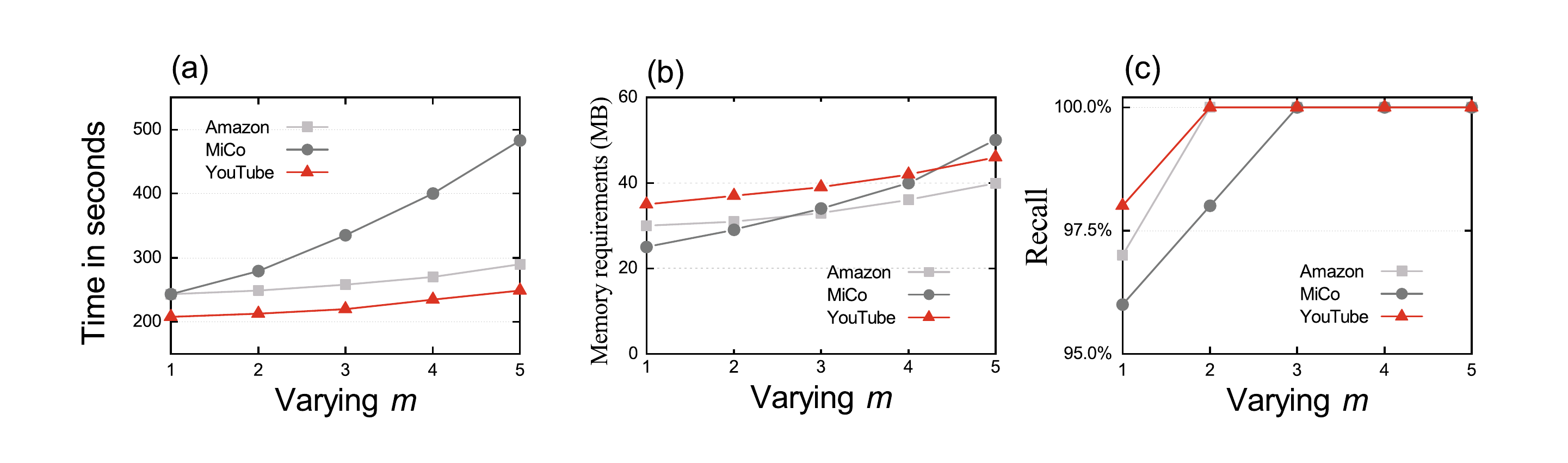}
    \vspace{-2ex}
    \caption{Exp-4: Influence of $m$}
    \label{fig:exp4}
\end{figure}

\stitle{Exp-4: Influence of $m$}. Fixing $k$ = $100$ and $\theta$ = $1.8K$, $2.9K$ and $0.29K$ for {\em Amazon}, {\em Mico} and {\em Youtube}, respectively, we varied $m$ from $1$ to $5$ in $1$ increments to test its influence \wrt efficiency, memory cost and recall for \apm. As shown in Figures~\ref{fig:exp4}(a)-\ref{fig:exp4}(c), 
(1) with the increase of $m$, the time overhead and memory cost of \apm grow up as well;
(2) on {\em Amazon} and {\em Youtube} (resp. {\em Mico}), when $m$ reaches 2 (resp. 3), the improvement on recall becomes insignificant, since recall values already approach 100\%. Hence, it is more appropriate to set $m$ as $2$ or $3$ on real life graphs.

\vspace{-2ex}

\eat{
\begin{figure*}[tb!]
\begin{center}
\subfigure[Response time({\em Mico})]{\label{rt-opt}
{\includegraphics[width=4.cm,height=2.35cm]{fig/RuntimeofOPT.eps}}} 
\subfigure[Recall ({\em Mico})]{\label{recall-rst}
{\includegraphics[width=4.cm,height=2.35cm]{fig/RecallofOPT.eps}}} \hfill
\end{center}
\vspace{-4ex}
\caption{Effects of optimization techniques} 
\label{fig-disrquery}
\vspace{-2ex}
\end{figure*}
}


\eat{\stitle{Exp-4: Optimizations}. 
On real-life graphs, we tested effectiveness of our optimization technique and find that the pruning strategy is very effectively, can achieve an improvement on efficiency of up to 2 orders of magnitude. Take {\em Mico} as example, \aprtopk only takes 48.8\% time of $\kw{AprTopK}_{nopt}$, on average. Due to space constraint, we omit details here. Interested readers may refer to~\cite{full} for more details. \looseness=-1}

\eat{
This experiment proves the impact of optimization and approximation strategies on frequent patterns mining. \nopt, both UplevelPruning and random pick strategies are disabled. \upruning, only UplevelPruning optimization enabled. \rpick, only random pick strategy enabled. A summary is illustrated in Figures~\ref{rt-opt}. \rpick has the best effective optimization, followed by \upruning. Through the combination of \rpick and \upruning, our \apm can achieve an improvement of up to 2 orders of magnitude.

This experiment shows the effect of our strategies to improve the recall on frequent patterns mining. \nrst, both secondary verification and make up strategies are disabled. \sverify, only secondary verification strategy enabled. \mk, only compensation strategy enabled. As shown in Figures~\ref{recall-rst}, (1) \sverify is the most effective way to improve the recall. \mk only has an effect at a small threshold. (2) By adopting \sverify and \mk strategies, \apm can increase the recall by 10\%.
}


\section{Conclusion}
\label{sec-conclusion}

\eat{
We have proposed a approximate technique ({\em travel embedding}) to identify top-$k$ patterns from large graphs. To do this, we have introduced the \kw{DI} model to calculate support of a pattern;the algorithm combines the strategy of ``look-ahead \& backtracking'' to discover frequent patterns. In particular, we have also developed an approximate algorithm with adjustable parameters({\em random pick}) to efficiently discover top-$k$ patterns.
Our experimental study has verified the efficiency, recall and scalability of the algorithm. We hence contend that our approach yields a promising tool for big graph analysis. \looseness=-1
}
In this paper, we developed an approach to mining {\em near optimal} top-$k$ patterns. We first formalize the \topkpm problem by incorporating viable metrics to measure support and interestingness of patterns. 
We then develop an algorithm \aprtopk to identify top-$k$ patterns efficiently and accurately. The algorithm applies a ``level-wise'' strategy, which ensures {\em early termination property}, to discover top-ranked patterns that are not only frequent but also interesting. 
To facilitate support evaluation, we devised a technique to compute the lower bound of support with smart traverse strategy and compact data structure. 
Our experimental study has verified the efficiency, memory footprint, recall and scalability of our algorithm. We hence contend that our approach yields a promising tool for big graph analysis. \looseness=-1

The study of \topkpm is still in its infancy. 
One direction concerns pruning technique that may lead to the decrease of costs (computational and space costs) without sacrificing recall. 
Metrics for measuring importance of patterns also need to investigate. 
Another interesting topic is to identify top-$k$ patterns with different matching semantics, \eg graph simulation, inexact matching, etc. 
It is also worth extending \aprtopk under distributed scenario, to leverage parallel computation. \looseness=-1

\section*{CRediT authorship contribution statement} 
Xin Wang: Conceptualization, Methodology, Formal analysis, Writing – original draft, Writing – review \& editing. Zhuo Lan: Methodology, Software, Writing – original draft, Visualization. Yu-Ang He: Software, Validation, Writing – original draft, Visualization. Yang Wang: Methodology, Supervision. Zhi-Gui Liu: Methodology, review \& editing. Wen-Bo Xie: Conceptualization, Formal analysis, Supervision, Writing – review \& editing.

\section*{Acknowledgments}
This work is supported by National Natural Science Foundation of China [grant number 62172102], and National Key Research and Development Program of China [grant number 2017YFA0700800], and Young Scholars Development Fund of SWPU [grant number 202199010142].

\balance
\bibliographystyle{apacite}

\bibliography{mybib}

\end{document}